\documentclass[natbib,12pt,preprint]{aastex}
\usepackage{psfig,lscape}
\def\lap{\lower.5ex\hbox{$\; \buildrel < \over \sim \;$}}
\def\gap{\lower.5ex\hbox{$\; \buildrel > \over \sim \;$}}

\begin{document}
\title{Imaging and Demography of the Host Galaxies of High-Redshift
Type Ia Supernovae\altaffilmark{1,2}}
\vspace{-0.5in} 

\author{ Benjamin F. Williams,\altaffilmark{3} Craig
J. Hogan,\altaffilmark{4} Brian Barris,\altaffilmark{5} Pablo
Candia,\altaffilmark{6} Peter Challis,\altaffilmark{3} Alejandro
Clocchiatti,\altaffilmark{6} Alison L. Coil,\altaffilmark{7} Alexei
V. Filippenko,\altaffilmark{7} Peter Garnavich,\altaffilmark{8} Robert
P. Kirshner,\altaffilmark{3} Stephen T. Holland,\altaffilmark{8}
Saurabh Jha,\altaffilmark{3,7} Kevin Krisciunas,\altaffilmark{6,9}
Bruno Leibundgut,\altaffilmark{10} Weidong Li,\altaffilmark{7} Thomas
Matheson,\altaffilmark{3} Jose Maza,\altaffilmark{11} Mark
M. Phillips,\altaffilmark{9} Adam G. Riess,\altaffilmark{12} Brian
P. Schmidt,\altaffilmark{13} Robert
A. Schommer,\altaffilmark{14\dagger} R. Chris Smith,\altaffilmark{6}
Jesper Sollerman,\altaffilmark{15} Jason Spyromilio,\altaffilmark{10}
Christopher Stubbs,\altaffilmark{3} Nicholas
B. Suntzeff,\altaffilmark{6} and John L. Tonry\altaffilmark{5}}

\altaffiltext{1}{{\tiny Based on observations with the NASA/ESA Hubble Space
Telescope obtained at the Space Telescope Science Institute, which is
operated by the Association of Universities for Research in Astronomy,
Inc., under NASA contract NAS5-26555.}}
\altaffiltext{2}{{\tiny Based on observations obtained at the Cerro-Tololo
Inter-American Observatory, which is operated by the Association of
Universities for Research in Astronomy, Inc., under NASA contract
NAS5-26555.}}
\altaffiltext{3}{{\tiny Harvard-Smithsonian Center for Astrophysics, 60 Garden Street, Cambridge, MA 02138.  \hbox{williams@head-cfa.harvard.edu}; \hbox{challis@cfa.harvard.edu}; \hbox{kirshner@cfa.harvard.edu}; \hbox{tmatheson@cfa.harvard.edu}; \hbox{cstubbs@cfa.harvard.edu}}}
\altaffiltext{4}{{\tiny University of Washington, Astronomy Dept., Box 351580, Seattle, WA  98195-1580. \hbox{hogan@astro.washington.edu}}}
\altaffiltext{5}{{\tiny Institute for Astronomy, University of Hawaii, Manoa, Honolulu, HI  96822.  \hbox{jt@ifa.hawaii.edu}; \hbox{barris@ifa.hawaii.edu}}}
\altaffiltext{6}{{\tiny Cerro Tololo Inter-American Observatory, Casilla 603,
La Serena, Chile; 
\hbox{aclocchi@astro.puc.cl}; \hbox{csmith@ctios2.ctio.noao.edu}; \hbox{nick@ctiow5.ctio.noao.edu}; \hbox{pcandia@ctio.noao.edu}; \hbox{kevin@ctiosz.ctio.noao.edu}}}
\altaffiltext{7}{{\tiny Department of Astronomy, University of California, Berkeley, CA 94720-3411. \hbox{alex@astron.berkeley.edu};
\hbox{acoil@astron.berkeley.edu}; \hbox{sjha@astron.berkeley.edu};
\hbox{weidong@astron.berkeley.edu}}}
\altaffiltext{8}{{\tiny University of Notre Dame, Department of Physics, 225
Nieuwland Science Hall, Notre Dame, IN
46556. \hbox{pgarnavich@miranda.phys.nd.edu}; \hbox{sholland@nd.edu}}}
\altaffiltext{9}{{\tiny Las Campanas Observatory, Casilla 601, La Serena, Chile. \hbox{mmp@lco.cl}}}
\altaffiltext{10}{{\tiny European Southern Observatory, Karl-Schwarzschild-Strasse 2, D-85748 Garching, Germany. \hbox{bleibund@eso.org}; \hbox{jspyromi@eso.org}}}
\altaffiltext{11}{{\tiny Departamento d'Astronomia, Universidad de Chile, Casilla 26-D, Santiago, Chile. \hbox{jose@das.uchile.cl}}}
\altaffiltext{12}{{\tiny Space Telescope Science Institute, 3700 San Martin Drive, Baltimore, MD 21218. \hbox{ariess@stsci.edu}}}
\altaffiltext{13}{{\tiny The Research School of Astronomy and Astrophysics, the Australian National University, Weston Creek, ACT 2611 Australia.  \hbox{brian@mso.anu.edu.au}}}
\altaffiltext{14}{$\dagger${\tiny  Deceased.}}
\altaffiltext{15}{{\tiny Stockholm Observatory, AlbaNova, SE-106 91 Stockholm, Sweden. \hbox{jesper@astro.su.se}}\\\\\\\\}

\newpage

\begin{abstract}
We present the results of a study of the host galaxies of high
redshift Type Ia supernovae (SNe Ia).  We provide a catalog of 18
hosts of SNe Ia observed with the {\it Hubble Space Telescope (HST)}
by the High-$z$ Supernova Search Team (HZT), including images,
scale-lengths, measurements of integrated (Hubble equivalent) {\it
BVRIZ} photometry in bands where the galaxies are brighter than $m
\approx 25$ mag, and galactocentric distances of the supernovae.  
We compare the residuals of SN~Ia distance measurements from cosmological
fits to measurable properties of the supernova host galaxies that
might be expected to correlate with variable properties of the
progenitor population, such as host galaxy color and position of the
supernova.  We find mostly null results; the current data are
generally consistent with no correlations of the distance residuals
with host galaxy properties in the redshift range $0.42 < z < 1.06$.
Although a subsample of SN hosts shows a formally significant
(3$\sigma$) correlation between apparent $V-R$ host color and
distance residuals, the correlation is not consistent with the null
results from other host colors probed by our largest samples.  There
is also evidence for the same correlations between SN Ia properties
and host type at low redshift and high redshift.  These similarities
support the current practice of extrapolating properties of the nearby
population to high redshifts pending more robust detections of any
correlations between distance residuals from cosmological fits and
host properties.

\end{abstract}

\section{Introduction}

The claimed discovery \citep{riess1998,perlmutter1999} of
the acceleration of the expansion of the Universe was originally based
on the Hubble diagram of Type Ia supernovae (SNe Ia;
\citealp{schmidt1998,garnavich1998,perlmutter1997,perlmutter1998}).
Diagnosed as Type Ia by the lack of hydrogen and increased Si II
absorption in their spectra (see \citealp{filippenko1997} for a
review), SNe Ia appear to belong to a largely one-parameter family, where
differences in their intrinsic luminosities are correlated with
differences in their light-curve decline rates
\citep{phillips1993,hamuy1996d,hamuy1996a,hamuy1996b,hamuy1996c,riess1996}.
This relation was empirically determined from the Hubble diagram of a
large sample of nearby SNe Ia, but is not well understood
theoretically.  For example, the relation may be affected by different
chemical compositions (C/O ratios in the progenitor white dwarfs,
e.g., \citealp{hoflich1998,umeda1999b,umeda1999a}; $^{56}$Ni content
in the explosions, e.g., \citealp{pinto2000,mazzali2001}; main-sequence
mass and metallicity of progenitor, e.g., \citealp{dom2001}).  This
lack of solid theoretical understanding inspires questioning whether
there may be environmental and evolutionary trends of SNe Ia that
could propagate into their distance estimates.  The impact of these
distance measurements on cosmological models requires that any and all
possible indications of systematic trends of SN Ia properties with
redshift be checked.

The morphologies of high-redshift galaxies differ significantly from
those of low-redshift galaxies.  The spiral arms are less developed
and more chaotic \citep{abraham2001}, and the fraction of irregular
galaxies increases \citep{brinchmann1998,vandenbergh2001}.  The
possibility exists that these distant host galaxies have produced
progenitor populations leading to intrinsic luminosities of SNe
different from those seen in the nearby sample of SNe.  For example,
if the high-redshift hosts are in different phases of evolution from
the low-redshift hosts, they could contain dust with different
reddening laws \citep{totani1999}, or they could contain progenitor
stars of different abundance ratios \citep{hoflich2000,drell2000}.
Unfortunately, we cannot look at the stellar populations of the hosts
in detail because the individual stars cannot be resolved; however, we
can observe other host-galaxy properties, such as their integrated
colors, magnitudes, and the galactocentric distances (GCDs) of their
SNe.  These properties should correlate strongly with statistical
variation in progenitor population and thereby serve as statistical
proxies.  We can then compare these properties to residuals of the fit
of the SN Ia distances to the accelerating cosmological model.  If any
correlation is found in these comparisons, it will provide a hint that
conclusions about the accelerating universe, and the implied
cosmological constant, will require more sophisticated statistical
analysis incorporating such trends.  Conversely, a null result will
constrain models of such possible systematic effects.

Since the empirical relation which allows SNe Ia to be used as precise
distance indicators is not understood theoretically, the spread in
luminosities could be due to differing ages and/or chemical
compositions of the progenitors.  Many previous studies have noted that
the luminosities of SNe Ia are correlated with their distances from
the centers of their host galaxies and their host galaxy type.  In the
low-redshift sample of SNe Ia, events in elliptical galaxies occur at
larger GCDs and tend to be underluminous compared to events that
occur in spiral galaxies \citep{hamuy1996a,wang1997,ivanov2000}.  This
correlation suggests that the age of the SN Ia progenitor has an
effect on the peak brightness, because events hosted by ellipticals
likely come from older progenitors \citep{howell2001,hamuy2000}. The
correlation could also be explained by a metallicity effect; recently,
\citet{timmes2003} have shown that metallicity affects the amount of
$^{22}$Ne in the white dwarf, which affects the amount of Ni a SN Ia
explosion should make. Since ellipticals are super solar, they have
more $^{22}$Ne in their white dwarfs and less Ni produced. The
correlation also leads to a selection bias, as seen by
\citet{hamuy1999}.  The most distant SNe Ia in the low-redshift sample
tend to be those of the fainter variety, with higher GCD, hosted by
ellipticals.  \citet{hamuy2000} found evidence that the faintest
galaxies tend to host overluminous SNe Ia.

Fortunately, these correlations disappear when the estimated distances
(rather than luminosities) of the SNe are compared.  The effects of
the progenitor population on SN Ia luminosity in the low-redshift
sample are all accounted for through the decline rate vs. luminosity
relation ($\Delta m_{15}$ vs. $M$) without consideration for the host-galaxy
properties \citep{riess1999}.  The Hubble diagram of low-$z$ SNe Ia
displays no correlation of distance residuals with host population
indicators \citep{schmidt1998}.  Since the present-day stellar
populations include a range of stellar age and metallicity greater
than that spanned between the present and $z \approx 1$, this has been
one of the most powerful arguments to date that progenitor evolution
does not lead to a systematic bias in the high-$z$ Hubble diagram.

Further possible problems with the high-$z$ sample have also been
suggested.  \citet{drell2000} have found that incorporation of simple
models of SN Ia evolution allows many possible interpretations of the
high-redshift data, making it ``virtually impossible to pin down the
values of $\Omega_M$ and $\Omega_{\Lambda}$'' without an understanding
of the SN Ia process.  There is also the possibility that ``grey
dust'' in the intergalactic medium could be confused with a cosmic
acceleration \citep{aguirre1999}.  These effects appear unlikely to
greatly affect the Hubble diagram in light of the most recent
data \citep{tonry2003}, which suggest that SN Ia measurements are
consistent with a cosmological constant out to $z \approx 1$, where the
effects of a cosmological constant begin to diverge from those of a
systematic trend of SN~Ia properties with redshift.  Nevertheless,
the apparent differences between the high and low-redshift samples
highlight the need for further study of the possible differences
between the populations of SNe Ia at high and low redshift to see
whether there may be a smooth systematic trend with redshift that
could mimic a cosmological effect.


Recent tests for correlations between host-galaxy properties and SN~Ia
peak luminosities in the high-redshift sample have improved
constraints on the differences between the samples.  For example,
\citet{farrah2002} investigated 22 host galaxies at $z \approx 0.6$
observed by {\it HST}, finding the positions of the SNe to be in
conflict with the low extinction values measured for the events. These
studies did not show any correlations between host-galaxy type and SN
luminosity.  Most recently, \citet{sullivan2003} used the data from
the Supernova Cosmology Project, along with newly acquired host images
and spectra, to look for systematic differences between the high and
low-redshift SN~Ia samples.  Their high-redshift sample, uncorrected
for host reddening, suggested that SNe Ia hosted by late-type galaxies
have a larger intrinsic scatter than those found in early-type
galaxies, revealing the effects of dust in the high-$z$ sample.  On
the other hand, they measured a significant cosmological constant in
both the early and late-type samples, concluding that the measurement
is largely unaffected by host-galaxy dust.

In this paper, we study deep archival {\it HST} images of high-redshift 
SN~Ia host galaxies in order to look for correlations between their
properties and those of their SNe Ia.  The catalog presented here
comprises some of the highest quality imaging to date for a
statistical sample of high-$z$ SN hosts at a large range of redshifts.
Section 2 explains our data analysis technique, while \S 3 provides the
detailed results of our photometry and discusses the search for
correlations between the apparent photometric properties of the
host-galaxies and the residuals of the measured distances from smooth
cosmological Hubble diagrams.  Finally, \S 4 gives our
conclusions.

\section{Data Acquisition and Analysis}

The use of $HST$ to obtain light curves for several of the high-$z$ SN
Ia samples has created an archive of image data that allows the study
of the host galaxies' photometric properties at high resolution.  We
include all of the High-$z$ Supernova Search Team (HZT) proprietary and archival $HST$
data used for the determination of the SN~Ia light curves.  Table 1
provides a list of all of the data used for this study, including the
names of the SNe, the coordinates of the exposures, the filters used,
the number of epochs measured, the total exposure time in each filter,
and the total baseline covered by the data.  These data were not taken
with the idea that the different epochs would be combined to produce
deep images of the host galaxies.  The different epochs were taken at
different roll angles and with slight offsets, and there were SNe
contaminating the galaxy light.  Preparing the images for galaxy
photometry was therefore a significant challenge, as discussed below.

\subsection{Image Alignment and Stacking}

In most cases, the equatorial J2000 coordinate system (WCS) given to
the image by the $HST$ data pipeline was accurate enough to allow
pixel-to-pixel alignment by geometrically transforming each pixel to
the same physical position on a canvas with a previously determined
WCS.  In detail, we created blank images with WCS of one exposure
epoch; this was the reference coordinate system.  Then we determined
the geometric transformations, including translations and rotations,
necessary to match the pixel positions of the world coordinates of the
images taken at the other epochs to the pixel positions of the
reference world coordinates.  After these transformations were
accomplished, the images' physical coordinates were aligned as were
their world coordinates.

In some cases, the world coordinates of the images taken at different
epochs were not exactly aligned. In these cases, the transformed
images were misaligned by a few pixels, forcing an additional
translation of the images to align the centers of the few point
sources or compact galaxies in the images.  Once these final
adjustments were completed, the images were combined using the COMBINE
routine in IRAF\footnote{IRAF is distributed by the National Optical
Astronomy Observatory, which is operated by the Association of
Universities for Research in Astronomy, Inc., under cooperative
agreement with the National Science Foundation.}, allowing the
rejection of cosmic rays with the CRREJECT algorithm.  The final
galaxy images are shown in Figure 1.  These combined images provided
our deepest views of the host galaxies of the high-$z$ supernovae
observed by the HZT, comparable with many of the ground-based images
of hosts in the low-$z$ sample.  In a few cases, only galaxy template
images were taken with $HST$.  In these cases, the stacked images did
not go as deep, but there was no light contamination from the SN.  In
all other cases the images were still contaminated by the light from
the SNe. The images stacked from the multi-epoch supernova data have
filled arrows in Figure 1 indicating the location of the supernova event.

\subsection{Supernova Identification and Removal}

Once the exposures were combined to create the deepest image possible
in each case, the SNe and most probable hosts were identified using
coordinates and finder charts supplied by the HZT.  Because the SNe
were still bright during most of the epochs that the data were taken,
we adopted a procedure to remove the SNe with minimal impact on the
galaxy photometry. Since reference {\it HST} images were not available
for all of the hosts, we used a method that did not require a
reference host image.  Then we tested our method using the few hosts
for which we had reference data.  The most straightforward way to
remove the SN contamination was by linearly interpolating across the
SN in the final, combined images.  This routine gave us pixel-by-pixel
control over the area contaminated by the SN.

In four cases (SNe 1997ck, 2000dy, 2000ec, and 2000eh) no obvious
galaxy could be seen near the SN, and we could not be sure if a
compact galaxy was being completely overwhelmed by the SN light, or if
the SN was in a very low surface brightness host.  In three of these
cases (SNe 2000dy, 2000ec, and 2000eh), we were able to use
shallower template images of the galaxy taken at a later epoch by
$HST$.  For SN 1997ck no template images were taken and no host was
seen down to $m_{F850LP} \approx 27$ mag within 7$''$ of the event
after the SN was removed from the image.  In this case, there is a
good chance that the host was simply overwhelmed by the event.  For
SN 1999Q, the only galaxy detected close enough to the event to be the
host is 2.3$''$ south of the event.  For the events SNe 1999fj,
1999fk, and 1999fn the SN photometry was done from the ground;
however, shallow galaxy template images were obtained from $HST$.  In
these cases, the $HST$ template images, shown in Figure 1, were used
for all of the host galaxy studies.

An example of the images before and after SN removal is shown in
Figure 2.  This removal caused additional error in our photometry.  In
order to set limits on the systematic error introduced by our removal
process, we ran our photometry routines on template images of the
host galaxies taken long after the decline of the SNe.  These images
were obtained by $HST$ for use by the HZT to subtract the background
contribution from the SN light while obtaining the light curves for the
SNe.  The templates were not as deep as the combined epochs from the
light curves, and they were only made in all filters for a few of the
SN hosts.  We compared photometry done with the templates to that done
with the SN-removed image stacks (cf. Table 2).

Even though we had template and multi-epoch images in all 5 filters
for seven of the hosts, three of these hosts were, unfortunately,
completely overwhelmed by the light of the SNe during the multi-epoch
data.  We treated these three cases (SNe 2000dy, 2000ec, and 2000eh)
like all of the others, and the comparisons between the photometry
extracted from the stacked epochs and the template images are given in
Table 2.  Because these were some of the faintest galaxies in the
sample and the SN light was so dominant in the multi-epoch images,
these comparisons mark the worst-case scenario for our SN removal
method.  In fact, these cases were so severe that we used the
photometry from the shallower, template images for the catalog.  On
the other hand, the differences between the measurements taken using
both methods for the other 4 galaxies (SNe 2000dz, 2000ee, 2000eg,
and 2000ea) are shown in Table 2.  In these cases, the light from
the galaxies is not overwhelmed by the SNe.  Most of the hosts in the
sample, as shown in Figure 1, share this characteristic.  Therefore,
the effects of the SN removal technique for these four cases most
likely mimic the effects of the technique on the rest of the sample.
These tests revealed that our technique was causing no systematic
offsets in the galaxies' magnitudes in cases where the galaxy was
brighter than the SN in the multi-epoch data.  The most useful case
was that of SN 2000dz, which had significant, but not dominant, SN
contamination inside of one scale length.  The photometry for this
galaxy came out consistent within 1.5$\sigma$ for both the stacked,
modified image and the template image obtained after the fading of the
SN event.

\subsection{Photometry}

We performed aperture photometry for the host galaxies at 12 radii,
from $0.2''$ to $4.0''$ using the IRAF package APPHOT.  We measured
the surface brightness fluctuations in the local background on each of
these size scales in order to find the precision with which the
photometry could be determined.  The mean background level was
subtracted from each pixel.  The background-subtracted total counts in
each aperture were transformed into instrumental magnitudes for each
filter used by applying the zero-points of \citet{holtzman1995}.
Since these galaxies are at high redshift, and their spectral energy
distributions are not well constrained, we left these raw Space
Telescope (ST) magnitudes alone for the purposes of this study, as any
attempts to transform them to more standard Johnson-Cousins magnitudes
or to rest-frame magnitudes would have added unknown systematic errors
to our analysis.  Instead, we plot in Figure 3 the galaxy colors
versus redshift.  The plot appears to show that higher redshift hosts
are redder, but this correlation is most likely due to the systematic
effects caused by the lack of imposing K-corrections on our data.

Surface brightness profiles were also determined using the counts in
circular annuli.  These profiles were used to determine the galaxy
scale length and apparent magnitude.  Total galaxy magnitudes were
measured out to one scale length, the radius where the mean F814W
surface brightness dropped by a factor of $e$ from the central surface
brightness.  The errors on these scale lengths were calculated from
the measured radial surface brightness gradient of the galaxy and the
surface brightness errors from fluctuations in the surrounding sky
brightness.  Since the radial surface brightness often fell steeply,
more than one standard deviation in one pixel, we were often able to
calculate the scale length to a precision of less than one pixel.

\section{Discussion}

Table 3 provides a catalog of data for each galaxy, including the
radii of the galaxies used to measure their integrated magnitudes, the
projected distances from the galaxies' centers to their SNe,
the integrated magnitudes measured in all filters observed, the galaxy
redshift, the distance moduli ($m-M$) determined from the SN
light curves, and the $m-M$ determined from the galaxy redshift assuming
an $\Omega_{M} = 0.3$ and $\Omega_{\Lambda} = 0.7$ universe.  The
calculations necessary to acquire these numbers are described below.

\subsection{Projected Galactocentric Distances}

The SNe and their hosts were all imaged on the WF3 chip, which has an
angular pixel scale of 0.1 arcsec/pixel.  We measured the pixel
positions of the galaxies' centers of light and the centers of the SN
point-spread functions (PSFs) 
using the IRAF task IMEXAM where there was a well-defined center.
In cases where the galaxy was too extended for IMEXAM to find a
reliable galaxy center, we estimated the center position by eye.
These measurements were used to calculate the GCDs in units of
host-galaxy radii.  Environmental parameters within a galaxy change as
a function of distance from the galactic center, and population
properties change more quickly in a more compact galaxy than in a less
compact one.  A consistent way to compare the different GCDs of the
events is to normalize them to galaxy size.

\subsection{Distances}

The homogeneous sample of luminosity distances for the SNe~Ia was
obtained from their light curves by the HZT (\citealp{tonry2003}, and
references therein).  For our tests, we use the final distances
measured by the HZT, including their absorption corrections.  If there
are no other systematics affecting the high-$z$ sample
(e.g. progenitor population, chemical evolution), then we do not
expect to find any correlation between any parameter and distance
residual.  We therefore are not searching for correlations we would
expect; instead we are searching for correlations we would {\it not}
expect.  The distance moduli shown in Table 3 were obtained from the
log($H_0d$) distances in Table 15 of \citet{tonry2003} using the
formula
$$
m-M = 5 (\log (H_0d)_{tonry} - \log (65 ~km~s^{-1}~Mpc^{-1})) + 25.
$$ 
In this equation, the $\log(H_0d)_{tonry}$ is the distance value given
in Table 15 of \citet{tonry2003} and $m-M$ is the distance modulus in
Table 3, assuming a value of $H_0~=~65~km~s^{-1}~Mpc^{-1}$.  The
distance errors $\sigma_{m-M}$ in Table 3 were obtained using
$\sigma_{m-M} = 5 \sigma_{log(H_0d)}$, where $\sigma_{log(H_0d)}$ is
the error value in Table 15 of \citet{tonry2003}.  No distance has
been measured for SN 2000dy because this event was later determined not
to be a SN~Ia; however, this event was useful to study because it
provided a test of our analysis technique, discussed in \S 2.2.
Two other HZT events (SN 1999fo and SN 1999fu) were also later
determined not to be SNe~Ia, and they have been removed from this
study entirely.

In order
to look for correlations between the host-galaxy properties and the
discrepancy between these distance measures and those predicted from
theory, we calculated the host distances predicted from their
redshifts using a ``flat-lambda'' model \citep{carroll1992}, where the
luminosity distance, $D_L$, is given by
$$
D_L = c(1+z)/H_0 \int_{0}^{Z}((1 + z)^2 (1 + \Omega_M z) - z (2 + z) 
\Omega_{\Lambda})^{-0.5} dz,
$$ where $c$ is the speed of light and $z$ is the
redshift of the galaxy.  For the analysis below, we adopt a reference
model with $\Omega_M = 0.3$, $\Omega_{\Lambda} = 0.7$, consistent with
the concordance of other current databases (e.g., WMAP;
\citealp{spergel2003}).  Finally, the distance residuals used in our
analysis were calculated by subtracting the redshift distance from the
distance measured by the SN photometry.

The point should be made that for the distance-residual demographics,
the exact choice of cosmological parameters ($H_0$, $\Omega_M$,
$\Omega_{\Lambda}$) hardly matters to first order.  The measurement of
the cosmological constant is based on comparing the low-redshift
distances to the high-redshift distances.  Although we have used the
concordance values for these parameters, the high-redshift data alone
are not sensitive to these values, and therefore neither are the
distance residuals from the Hubble fit.  This insensitivity is
demonstrated in Figure 4, where we show the distance residuals for a
variety of different flat universe models, ranging from
$\Omega_{\Lambda} = 0$ to $\Omega_{\Lambda} = 1.0$.  In each case we
set the Hubble constant to minimize the sum of the squares of the
residuals.  The distance errors are shown for the $\Omega_{\Lambda} =
0.6$, $\Omega_{M} = 0.4$ case to compare the errors with the spread
over the possible cosmology choices.  The spread in residuals for
various models for any individual galaxy is always significantly
smaller than the distance errors for the galaxy, even covering this
large amount of parameter space.  Figure 5 shows the Hubble fit for
our sample and our adopted cosmology.  Events with early-type hosts
are marked by open squares, those with late-type hosts by closed
squares, and those with no host by open stars.

\subsection{Galaxy Classification}

High-redshift galaxy classification has long been difficult because
high-redshift galaxies often do not share the same kinds of
morphological properties as the low-redshift galaxies used to create
the classification system
(e.g., \citealp{abraham2001,vandenbergh2001}).  Our sample is no
different, with host galaxies like those of SNe 1997ce, 1998M, and 1999fj
having confused, irregular, and merging morphologies which look quite
different from those of the nearby sample.

Therefore, in classifying our sample of galaxies, we thought any
classifications beyond simply early and late types would be rather
arbitrary and could lead to confusion.  It was important, however, to
break down the sample into at least these two types of galaxies in
order to draw comparisons to previous studies of low and high-redshift
hosts.  These broad classifications are used because of the
differences in galaxy morphologies at high redshift, not because of
the quality of our images.  All hosts were imaged through the F814W or
F850LP filters.  These images probe the rest-frame $B$ or $V$,
revealing the distribution of young stars and providing reliable
morphology information.  Our subjective criteria were simple.
Galaxies which showed circular symmetry and lacking a disk were
classified as early, and all others (merger remnants, spirals, and
irregulars) were classified as late.  Table 3 gives our
classifications for each galaxy.

The morphologies of the hosts of 10 supernovae (SNe 1997ce, 1997cj, 1998J,
1998M, 1999U, 2000dz, 2000ea, 2000ec, 2000ee, 2000eg) were examined
independently by \citet{farrah2002}.  We classify the host of SN 1997cj 
as a late galaxy because of the appearance of an arm just above the bulge
in our image; \citet{farrah2002} classify this galaxy as early. We
classify the host of SN 1998M as a late galaxy because of the irregular appearance;
\citet{farrah2002} classify this galaxy as early.  
While these examples of disagreement underscore the subjective nature
of even these rough classifications, the rest of our classifications
agree.  In fact, the relative numbers of early and late galaxies are
the same in both of our studies.

With the small number of galaxies of each type in our sample (5 early
and 12 late), we could not draw any strong conclusions about the
differences between the two populations of SNe Ia.  In our sample, the
distance residuals of the SNe Ia in the early-type galaxies from the Hubble
fit are slightly larger than those of the SNe Ia in late galaxies,
unlike the results of the SN Cosmology Project (SCP;
\citealp{sullivan2003}), but the difference
is not statistically significant. \citet{sullivan2003} found that the
scatter of the distance residuals was larger for SNe Ia with late-type
hosts: their measured dispersion was 0.159 mag in early-type hosts
and 0.235 mag in late-type hosts.  This result is to be expected as
late-type galaxies tend to contain more dust, increasing the
uncertainty in converting apparent magnitudes to absolute magnitudes.
With our smaller sample, we calculate a dispersion for the late-type
hosts of 0.30 mag (12 galaxies), similar to the value measured by
\citet{sullivan2003}.  However, we find a dispersion in the early-type
hosts of 0.37 mag (5 galaxies), larger than that seen in the SCP
sample, but this difference is insignificant given our small sample
size.  If the distance residuals are converted to units of $\sigma$ by
dividing by the measured distance error, and the mean dispersion is
calculated, values of 1.42$\sigma$ and 0.98$\sigma$ are found for
the early and late-type samples, respectively.  Figure 6 shows a
histogram of the distance residuals for the early and late host
populations.  One possible reason for the large scatter may be the
HZT absorption corrections, which could introduce additional
uncertainty.  At the same time, the large spread in early types is
consistent with statistical expectation if there is no intrinsic
difference between the early and late-type samples, suggesting that
the HZT absorption corrections are accurate and recover the correct
zero-point.

In contrast, the GCD distribution of the events in the early-type
galaxies appears to be more skewed to larger radii than in the
late-type galaxies, in agreement with the trend seen in low-redshift
samples. \citet{ivanov2000} found that late-type hosts dominated the
low-redshift SN Ia events with deprojected GCD $<$7.5 kpc, and
early-type hosts dominated those events with deprojected GCD $>$7.5
kpc.  The distribution of GCDs in our sample is shown in Figure 7.
With such a small sample, we can note only coarse measures of the
differences between the populations.  Zero out of 12 SNe in late-type hosts
lie outside of 5 host radii, whereas 2 of the 5 SNe in early-type hosts do.  Two
out of 12 late types lie outside of 8 kpc, whereas 2 out of 5 early
types do.  The mean GCD of the early hosts is 2.8 and that of the late
hosts is 1.9 in units of scale radii.  Conversion of the GCDs to units
of kpc yields a similar result.  The mean GCDs of the events are 8.0
and 5.6 kpc for early and late-types, respectively.  The rms scatter
of the galactocentric distances is 3.3 host radii in early types, and
it is 1.4 host radii in late types.  When calculated in kpc, these
values are 8.3 and 4.5 for the early and late samples, respectively.
In the low-redshift sample early-type galaxies tend to host events
with higher GCDs \citep{hamuy1996a,wang1997,ivanov2000}.  Although the
scatter is large, the higher mean GCD measured for ellipticals in this
sample is consistent with the trend found in the low-redshift sample.

Finally, we find, in agreement with \citet{farrah2002}, a similar
ratio of the numbers of events in early and late-type galaxies as that
seen in the low-redshift sample (e.g., 23/62; \citealp{ivanov2000}),
with about one third of the events taking place in early-type
galaxies.  If extinction of events in spiral hosts were high owing to
dust, we should expect the high-$z$ samples to be biased against events
in late-type hosts.  The similarity in numbers of events in both types
of galaxies
is consistent with previous findings that the effects of extinction in
spirals do not significantly hinder the SN~Ia surveys
\citep{hatano1998,sullivan2003}.  In addition, the similarity between
the ratios of host types in the low and high-redshift samples suggests
that the same trends between SN Ia properties and host type in the
low-redshift sample are going to be seen in the high-redshift sample;
larger numbers of events will make this much clearer.

\subsection{Looking for Correlations:  Plots of Host-Galaxy Properties}

In preparing this sample of host galaxies, our aim was to search for
correlations between host galaxy properties and residuals of the fit
to the Hubble diagram.  Objective tests were performed between each
host property that we measured and the properties of the SNe to look
for empirical correlations that may not exist in the low-redshift
sample or may not be well understood theoretically.  If no new
correlations are found, the case for using SNe Ia as standard candles
will be strengthened.  On the other hand, any new correlations will
need to be explained and further tested to assess their affect on the
measurement of the cosmological constant.

Figures 8-12 show plots correlating the host-galaxy properties and the
properties of the SNe. Figure 8 shows the relation between the
distance residuals and the normalized angular GCD of the SNe in units
of host radii.  Figure 9 tests for a correlation between the
galaxies' apparent colors and the distance residuals, where each panel
provides a test for an observed color.  Figure 10 shows distance
residuals vs. host magnitude in the filters observed.  We have
subtracted the distance modulus from the hosts' apparent magnitudes.
Each panel provides a test for a different filter, except the upper
left, which combines the $B$ and $V$ results.  The Figure 11 plot
shows the galactocentric distances vs. the integrated apparent colors,
where each panel provides a test for a different color.  Finally,
Figure 12 investigates whether the visual extinction of the events is
influenced by the GCD.

We performed $\chi^2/dof$ fits to the plots shown in Figures 8-12.
These $\chi^2/dof$ fits allowed quantitative constraints to be placed
on the possibility of correlations between the properties of the host
galaxies and the residuals to the cosmological fits.  The results of
the fits are given in Table 4, where the first column is the parameter
tested on the abscissa, and the second column is the measurement that may
be correlated to the parameter.  The third column gives the number of
degrees of freedom in the fits, and the slope of the best-fitting line is
then given with errors in column 4.  The error values provided contain
all linear fits with $\chi^2/dof$ consistent with at least a 1\%
chance of matching the distribution.  Any value outside of this range
is therefore unacceptable at the 99\% confidence level.  A value of
$n/a$ is given where no reasonable straight-line fit was obtained.
The high $\chi^2/dof$ values for these parameters reveal that the
scatter is too large for some subsamples, probably indicating that
some SN Ia distance errors have been underestimated.

A slope of zero is acceptable at the 99\% confidence level for all but
two of the linear fits (cf. Table 4), suggesting no correlation. The
only parameters which show possible correlations with distance
residual are those of $B-V$ and $V-R$ color.  The best $\chi^2/dof$
fits to the residual vs. $B-V$ plot and the residual vs. $V-R$ plot
have slopes of $-0.51$ and $-0.83$, respectively.  Slopes of 0 for these
parameters are ruled out at 99\% confidence.  This subsample of hosts
are all at similar redshifts ($0.42 < z < 0.54$), so that the lack of
K-corrections is unlikely to explain the scatter seen here.  On the
other hand, with only 5 and 6 data points, these detections are not
robust.  For example, the $B-V$ data points, while being best fit by a
line with slope $-0.51$, has a Spearman rank coefficient of 0.1,
suggesting no correlation.  The $V-R$ correlation is more severe and
shows a consistent Spearman rank coefficient; however, neither of
these correlations is consistent with the tests performed on the
largest samples in $R-I$ and $I-Z$.  All of the galaxies with $B-V$
and $V-R$ measurements are from the same subsample, events measured in
the year 2000.  Further suspicion on this correlation is cast by the
large $\chi^2/dof$ (i.e., goodness of fit) for the small subsample
(cf. Fig. 9).

While a possible correlation is seen between distance residuals and
apparent host colors in the year 2000 subsample, this subsample of
distances shows significantly greater scatter in the Hubble fit than
the rest of the distances used in this study.  The $\chi^2/dof$ values
for the Hubble fits to the SN Ia distances measured before the year
2000 and distances measured in the year 2000 are 0.9 and 3.8,
respectively.  The large $\chi^2/dof$ value for the 2000 events are also
responsible for the large $\chi^2/dof$ values for the plots of
distance residual vs. $M_B$ and $M_V$ because those plots contain
exclusively year 2000 events.  It is of concern that the results from
these events are also responsible for the correlation detections
described above.  The data set for these events is comprised of 5-filter 
photometry, allowing a detailed analysis on dust and SN colors
which is in progress and will provide more discussion on this issue
\citep{jha2004}.

Inspired by all of the interesting results regarding the extinction
measures ($A_V$) for SNe~Ia from previous studies (see \S 1),
we checked for a correlation between the GCD of the events and the
measured extinction.  Figure 12 shows a plot of these measurements
using the extinction values of \citet{tonry2003}.  These values are
measured by simultaneously and iteratively fitting the observed colors
to a reddening-free set of synthetic light curves.  The extinction
measurements are all assigned errors of 0.1 mag, as measured by
\citet{schmidt1998}.  The plot reveals no correlation; a least-squares
fit to the data gives a slope of 0.02.  A jack-knife test, where the
one obvious outlier is removed from the data set, yields a slope of
$-0.01$.  There is no evidence for any correlation between event GCD and
extinction at high $z$.  No significant difference is seen between
extinction values of the early and late samples, again suggesting that
extinction is not of great concern in late-type hosts, although it
should be pointed out that the 4 highest extinction events were in
late-type hosts and 2 of these events occur less than half of a scale
length out.

\section{Conclusions and Future Work}
 
We have supplied a catalog of high-quality images and measured the
photometric properties of 19 high-redshift SN candidate host-galaxies
(but one of these was not a true SN~Ia).  Simple tests show
hints of a correlation between host-galaxy apparent $B-V$ and $V-R$
color and SN~Ia distance determinations.  The scatter of the distance
measurements appears to exceed the measured errors for events studied
in the year 2000.  These hints are currently based on just a handful
of galaxies with large distance residuals, but they need to be further
investigated.  Although we have used the best currently available
distances to the Fall 2000 events, their light curves were fit using
$HST$ data alone and the calibration has not been exhaustively
verified.  Our results suggest that both the distances and the error
estimates on these points may be revised upon closer examination.
Such an examination is currently being performed by \citet{jha2004}.

We find trends between host type and location of the SNe, as well as
the relative numbers of SNe in different host types, in excellent
agreement with the low-$z$ sample.  The extinction measurements,
galactocentric distances, and host types for the events in our sample
are consistent with previous studies that suggest host extinction does
not have a strong effect on SN properties even for events in late-type
hosts.  These similarities support the current practice of
extrapolating properties of the nearby population to high redshifts
pending more robust detections of any correlations between distance
residuals from cosmological fits and host properties.  Further testing
will be required to determine if significant reduction in distance
error can be achieved using such demographic correlations.

Our catalog contains galaxy photometry that can be used for more
sophisticated analysis methods seeking systematic evolution of the SN
Ia population with redshift including models of the galaxy progenitor
population.  More accurate measurements of galaxy colors, including
accurate transformations to intrinsic colors, will allow more
stringent constraints to be placed on correlations between residuals
of the SN Ia distances to the Hubble fit and host galaxy color.

Support for this work at the University of Washington was provided by
NSF grant AST-009855 and by NASA grant AR-09201 from the Space
Telescope Science Institute (STScI), which is operated by the
Association of Universities for Research in Astronomy, Inc., under
NASA contract NAS5-26555.  A.C. aknowledges the support of CONICYT
(Chile) through FONDECYT grants 1000524 and 7000524.  A.V.F.'s group
at U.C. Berkeley is supported by NSF grant AST-0206329 and by NASA
grant GO-09118 from STScI.  We thank the anonymous referee for many
very helpful suggestions.


\clearpage

\begin{landscape}

\begin{deluxetable}{ccccccc}
\small
\tablewidth{8.0in}
\tablecaption{Data obtained from the $HST$ data archive.}
\tableheadfrac{0.05}
\tablehead{
\colhead{{SNe}} &
\colhead{{RA (2000)}} &
\colhead{{DEC (2000)}} &
\colhead{{Filter}}  &
\colhead{{Epochs}} &
\colhead{{Exp. (sec)}} & 
\colhead{{Baseline}}
}
\startdata
SN1997ce  &   17:07:48  &   44:00:39  &   F675W   &  7  &  6200  & 1997-05-15 - 1998-06-29\\
SN1997ce  &   17:07:48  &   44:00:39  &   F814W   &  7  &  8000  & 1997-05-15 - 1998-06-29\\
SN1997cj  &   12:37:10  &   62:26:01  &   F675W   &  6  &  4800  & 1997-05-25 - 1998-05-18\\
SN1997cj  &   12:37:10  &   62:26:01  &   F814W   &  6  &  6700  & 1997-05-25 - 1998-05-18\\
SN1997ck  &   16:53:00  &   35:02:59  &   F850LP   &  6  &  14000  & 1997-05-14 - 1997-06-21\\
SN1998I  &  08:04:55  &  05:16:01  &   F675W   &  6  &  4800  & 1998-02-02 - 1999-02-08\\
SN1998I  &  08:04:55  &  05:16:01  &   F814W   &  6  &  7400  & 1998-02-02 - 1999-02-08\\
SN1998J  &  09:31:13  &  -04:45:18  &   F675W   &  6  &  4800  & 1998-02-03 - 1999-02-14\\
SN1998J  &  09:31:13  &  -04:45:18  &   F814W   &  6  &  7000  & 1998-02-03 - 1999-02-14\\
SN1998M  &   11:33:47  &  04:04:48  &   F675W   &  5  &  4000  & 1998-02-03 - 1998-03-30\\
SN1998M  &   11:33:47  &  04:04:48  &   F814W   &  5  &  5800  & 1998-02-03 - 1998-03-30\\
SN1998aj  &  09:27:59  &  -05:00:02  &   F850LP   &  6  &  19600  & 1998-04-09 - 1998-10-16\\
SN1999Q  &  08:00:48  &  05:31:45  &   F675W   &  6  &  4800  & 1999-02-01 - 1999-03-07\\
SN1999Q  &  08:00:48  &  05:31:45  &   F814W   &  6  &  7200  & 1999-02-01 - 1999-03-08\\
SN1999U  &  09:26:46  &  -05:37:39  &   F675W   &  6  &  4800  & 1999-02-01 - 1999-03-08\\
SN1999U  &  09:26:46  &  -05:37:39  &   F814W   &  6  &  6600  & 1999-02-01 - 1999-03-08\\
SN1999fj  &  02:28:20  &  00:39:08  &   F814W   &  1  &  2400  & 2000-09-21\\
SN1999fj  &  02:28:20  &  00:39:08  &   F850LP   &  1  &  2600  & 2000-09-21\\
SN1999fk  &  02:28:55  &  01:16:26  &   F814W   &  2  &  6600  & 2000-09-17 - 2001-02-24\\
SN1999fk  &  02:28:55  &  01:16:26  &   F850LP   &  2  &  6800  & 2000-09-17 - 2001-02-24\\
SN1999fn  &  04:14:07  &  04:17:52  &   F675W   &  1  &  2400  & 2000-04-03\\
SN1999fn  &  04:14:07  &  04:17:52  &   F814W   &  1  &  2400  & 2000-04-03\\
SN1999fn  &  04:14:07  &  04:17:52  &   F850LP   &  1  &  2600  & 2000-04-03\\
SN2000dy  &   23:25:37  &  -00:22:31  &   F555W   &  3  &  3000  & 2000-12-07 - 2001-05-05\\
SN2000dy  &   23:25:37  &  -00:22:31  &   F675W   &  4  &  5600  & 2000-12-07 - 2001-05-05\\
SN2000dy  &   23:25:37  &  -00:22:31  &   F814W   &  4  &  7800  & 2000-12-06 - 2001-05-05\\
SN2000dz  &   23:30:42  &  00:18:45  &   F450W   &  3  &  3000  & 2000-11-13 - 2001-05-10\\
SN2000dz  &   23:30:42  &  00:18:45  &   F555W   &  5  &  4400  & 2000-11-13 - 2001-05-10\\
SN2000dz  &   23:30:42  &  00:18:45  &   F675W   &  7  &  9600  & 2000-11-13 - 2001-05-10\\
SN2000dz  &   23:30:42  &  00:18:45  &   F814W   &  7  &  14900  & 2000-11-13 - 2001-05-10\\
SN2000dz  &   23:30:42  &  00:18:45  &   F850LP   &  3  &  7700  & 2000-11-13 - 2001-05-10\\
SN2000ea  &  02:09:55  &  -05:28:12  &   F450W   &  4  &  3600  & 2000-11-10 - 2001-06-08\\
SN2000ea  &  02:09:55  &  -05:28:12  &   F555W   &  6  &  5200  & 2000-11-10 - 2001-06-08\\
SN2000ea  &  02:09:55  &  -05:28:15  &   F675W   &  8  &  10400  & 2000-11-10 - 2001-07-01\\
SN2000ea  &  02:09:55  &  -05:28:15  &   F814W   &  8  &  16300  & 2000-11-10 - 2001-07-01\\
SN2000ea  &  02:09:55  &  -05:28:12  &   F850LP   &  5  &  10400  & 2000-11-10 - 2001-06-08\\
SN2000ec  &  02:11:33  &  -04:13:51  &   F450W   &  4  &  3800  & 2000-11-10 - 2001-06-21\\
SN2000ec  &  02:11:33  &  -04:13:51  &   F555W   &  6  &  5200  & 2000-11-10 - 2001-06-21\\
SN2000ec  &  02:11:33  &  -04:13:53  &   F675W   &  8  &  10400  & 2000-11-10 - 2001-06-21\\
SN2000ec  &  02:11:33  &  -04:13:53  &   F814W   &  8  &  16300  & 2000-11-10 - 2001-06-21\\
SN2000ec  &  02:11:33  &  -04:13:51  &   F850LP   &  4  &  10500  & 2000-11-10 - 2001-06-21\\
SN2000ee  &  02:27:35  &  01:11:55  &   F450W   &  4  &  3800  & 2000-11-12 - 2001-06-22\\
SN2000ee  &  02:27:35  &  01:11:55  &   F555W   &  6  &  5200  & 2000-11-11 - 2001-06-22\\
SN2000ee  &  02:27:35  &  01:11:55  &   F675W   &  7  &  8400  & 2000-11-11 - 2001-06-22\\
SN2000ee  &  02:27:35  &  01:11:55  &   F814W   &  7  &  12000  & 2000-11-11 - 2001-06-22\\
SN2000ee  &  02:27:35  &  01:11:55  &   F850LP   &  5  &  9900  & 2000-11-11 - 2001-06-22\\
SN2000eg  &  02:30:22  &  01:03:54  &   F450W   &  4  &  3800  & 2000-11-12 - 2001-06-20\\
SN2000eg  &  02:30:22  &  01:03:54  &   F555W   &  6  &  5200  & 2000-11-12 - 2001-06-20\\
SN2000eg  &  02:30:22  &  01:03:54  &   F675W   &  7  &  8400  & 2000-11-12 - 2001-06-20\\
SN2000eg  &  02:30:22  &  01:03:54  &   F814W   &  7  &  12000  & 2000-11-12 - 2001-06-20\\
SN2000eg  &  02:30:22  &  01:03:54  &   F850LP   &  4  &  9900  & 2000-11-12 - 2001-06-20\\
SN2000eh  &  04:15:02  &  04:23:27  &   F450W   &  3  &  3000  & 2000-11-15 - 2001-07-19\\
SN2000eh  &  04:15:03  &  04:23:24  &   F555W   &  5  &  4000  & 2000-11-15 - 2001-07-19\\
SN2000eh  &  04:15:03  &  04:23:23  &   F675W   &  6  &  8100  & 2000-11-15 - 2001-07-19\\
SN2000eh  &  04:15:03  &  04:23:23  &   F814W   &  6  &  10500  & 2000-11-15 - 2001-07-19\\
SN2000eh  &  04:15:02  &  04:23:27  &   F850LP   &  3  &  6900  & 2000-11-15 - 2001-07-20\\
\enddata
\end{deluxetable}
\end{landscape}

\clearpage

\begin{deluxetable}{ccccccc}
\small
\tablewidth{6.5in}
\tablecaption{Residuals of galaxy photometry.\tablenotemark{a}}
\tableheadfrac{0.05}
\tablehead{
\colhead{{SN}} &
\colhead{{Rad. ($''$)}\tablenotemark{b}} &
\colhead{{$B (mag$)}} &
\colhead{{$V (mag$)}} &
\colhead{{$R (mag$)}}  &
\colhead{{$I (mag$)}} &
\colhead{{$Z (mag$)}} 
}
\tablenotetext{a}{Galaxy photometry performed on stacked
epochs with SNe removed by linear interpolation ,and galaxy photometry
performed on template galaxy images taken after the SN had faded.}
\tablenotetext{b}{Scale length of the host galaxy.  Measurement technique described in text.}
\startdata
2000dz & -0.01$\pm$0.02 & 0.03$\pm$0.15 & 0.11$\pm$0.08 & 0.02$\pm$0.03 & 0.06$\pm$0.04 & 0.00$\pm$0.11\\
2000ee & -0.06$\pm$0.05 & 0.45$\pm$0.50 & 0.08$\pm$0.20 & -0.03$\pm$0.09 & 0.12$\pm$0.09 & 0.13$\pm$0.21\\
2000eg & 0.00$\pm$0.04 & -0.26$\pm$0.15 & -0.08$\pm$0.05 & -0.04$\pm$0.05 & -0.02$\pm$0.07 & -0.01$\pm$0.12\\
2000ez & 0.01$\pm$0.04 & -0.18$\pm$0.22 & -0.04$\pm$0.15 & -0.02$\pm$0.06 & -0.03$\pm$0.06 & 0.05$\pm$0.12\\
2000dy & -0.20$\pm$0.17 & --- & 1.01$\pm$0.72 & 0.08$\pm$0.52 & 0.23$\pm$0.43 & ---\\
2000ec & -0.14$\pm$0.08 & --- & -0.69$\pm$0.72 & -0.10$\pm$0.31 & -0.19$\pm$0.21 & -0.16$\pm$0.65\\
2000eh & -0.26$\pm$3.57 & 1.68$\pm$0.82 & 0.05$\pm$0.49 & -0.03$\pm$0.31 & 0.09$\pm$0.32 & -0.13$\pm$2.12\\
\enddata
\end{deluxetable}

\clearpage

\thispagestyle{empty}
\topmargin 0.8in
\begin{landscape}
\begin{deluxetable}{cccccccccccc}
\tablewidth{10.2in}
\tablecaption{The host-galaxy photometry from the stacked $HST$ images.}
\tableheadfrac{0.05}
\small
\tablehead{
\colhead{{SNe}} &
\colhead{{GCD ($''$)}} &
\colhead{{Rad. ($''$)}} &
\colhead{{$m_{F450W}$}} &
\colhead{{$m_{F555W}$}}  &
\colhead{{$m_{F675W}$}} & 
\colhead{{$m_{F814W}$}} &
\colhead{{$m_{F850LP}$}}  &
\colhead{{Z}} & 
\colhead{{($m-M$)$_{SN}$}} & 
\colhead{{($m-M$)$_{Z}$}} & 
\colhead{{Type}}
}
\startdata
SN1997ce & 0.41 & 0.46$\pm$0.07 & -- & -- & 22.80$\pm$0.13 & 22.39$\pm$0.15 & -- & 0.44 & 42.08$\pm$0.21 & 42.13 & late\\
SN1997cj & 0.76 & 0.35$\pm$0.02 & -- & -- & 22.42$\pm$0.03 & 21.94$\pm$0.04 & -- & 0.50 & 42.55$\pm$0.23 & 42.46 & late\\
SN1997ck & -- & -- & -- & -- & -- & -- & -- & 0.97 & 44.13$\pm$0.38 & 44.21 & --\\
SN1998I & $\leq$0.10 & 0.64$\pm$0.12 & -- & -- & 23.32$\pm$0.31 & 22.30$\pm$0.23 & -- & 0.89 & 43.81$\pm$0.26 & 43.98 & late\\
SN1998J & 0.23 & 0.62$\pm$0.04 & -- & -- & 23.36$\pm$0.09 & 22.30$\pm$0.06 & -- & 0.83 & 44.01$\pm$0.26 & 43.80 & late\\
SN1998M & 2.39 & 1.29$\pm$0.27 & -- & -- & 20.94$\pm$0.13 & 20.51$\pm$0.19 & -- & 0.63 & 42.92$\pm$0.23 & 43.06 & late\\
SN1998aj & 1.69 & 0.32$\pm$0.73 & -- & -- & -- & -- & 23.35$\pm$0.15 & 0.83 & 44.43$\pm$0.35 & 43.80 & early\\
SN1999Q & 2.27 & 0.31$\pm$0.07 & -- & -- & 24.36$\pm$0.14 & 23.76$\pm$0.14 & -- & 0.46 & 42.62$\pm$0.21 & 42.24 & early\\
SN1999U & 0.41 & 0.59$\pm$0.07 & -- & -- & 23.16$\pm$0.08 & 22.73$\pm$0.10 & -- & 0.50 & 42.54$\pm$0.35 & 42.46 & early\\
SN1999fj & 0.95 & 0.52$\pm$0.17 & -- & -- & -- & 22.94$\pm$0.17 & 22.29$\pm$0.23 & 0.82 & 43.76$\pm$0.26 & 43.76 & late\\
SN1999fk & 1.32 & 0.29$\pm$0.02 & -- & -- & -- & 22.73$\pm$0.09 & 22.19$\pm$0.08 & 1.06 & 44.23$\pm$0.27 & 44.45 & late\\
SN1999fn & 0.32 & 0.33$\pm$0.97 & -- & -- & 25.63$\pm$0.59 & 24.78$\pm$0.38 & 25.16$\pm$0.94 & 0.49 & 42.18$\pm$0.20 & 42.40 & late\\
SN2000dy & 0.1 & 0.28$\pm$0.09 & -- & 25.95$\pm$0.65 & 24.56$\pm$0.40 & 23.99$\pm$0.32 & -- & 0.61 & -- & 42.97 & early\\
SN2000dz & 0.35 & 0.44$\pm$0.01 & 23.50$\pm$0.10 & 22.87$\pm$0.04 & 22.01$\pm$0.01 & 21.64$\pm$0.01 & 21.48$\pm$0.08 & 0.50 & 42.70$\pm$0.26 & 42.46 & late\\
SN2000ea & 1.49 & 0.36$\pm$0.01 & 24.45$\pm$0.14 & 23.87$\pm$0.07 & 22.90$\pm$0.03 & 22.38$\pm$0.02 & 21.99$\pm$0.05 & 0.42 & 41.39$\pm$0.27 & 42.01 & late\\
SN2000ec & 0.10 & 0.29$\pm$0.04 & -- & 25.14$\pm$0.32 & 24.84$\pm$0.23 & 24.17$\pm$0.15 & 23.83$\pm$0.43 & 0.47 & 42.64$\pm$0.21 & 42.30 & early\\
SN2000ee & 0.67 & 0.45$\pm$0.02 & 24.76$\pm$0.26 & 24.15$\pm$0.08 & 23.31$\pm$0.05 & 22.77$\pm$0.03 & 22.48$\pm$0.10 & 0.47 & 42.65$\pm$0.21 & 42.30 & late\\
SN2000eg & 0.96 & 0.28$\pm$0.01 & 24.22$\pm$0.11 & 23.10$\pm$0.03 & 21.75$\pm$0.01 & 21.01$\pm$0.01 & 20.59$\pm$0.01 & 0.54 & 42.12$\pm$0.24 & 42.66 & late\\
SN2000eh & $<$0.10 & 0.24$\pm$0.01 & 26.75$\pm$0.79 & 25.23$\pm$0.29 & 23.89$\pm$0.07 & 23.18$\pm$0.05 & 22.72$\pm$0.11 & 0.49 & 42.06$\pm$0.21 & 42.40 & early\\
\enddata
\end{deluxetable}
\end{landscape}
\clearpage

\topmargin 0.0in
\newpage
\clearpage

\thispagestyle{empty}
\topmargin 0.8in
\begin{deluxetable}{cccccc}
\tablewidth{7.0in}
\tablecaption{Linear fits of SNe Ia and host properties}
\tableheadfrac{0.05}
\small
\tablehead{
\colhead{{X axis}} &
\colhead{{Y axis}} &
\colhead{{d.o.f.}} &
\colhead{{Best Slope\tablenotemark{a}}} &
\colhead{{Best $\chi^2/dof$}} &
\colhead{{$\rho\tablenotemark{b}$}}
}
\startdata
GCD & Distance Residual (mag) & 16 & 0.02$\pm$0.05 & 1.836 & 0.0\\
$M_B$ & Distance Residual (mag) & 5 & $n/a$ & 3.677 & $-0.3$\\
$M_V$ & Distance Residual (mag) & 5 & $n/a$ & 3.716 & $-0.3$\\
$M_R$ & Distance Residual (mag) & 13 & 0.00$\pm$0.06 & 1.979 & 0.0\\
$M_I$ & Distance Residual (mag) & 15 & 0.01$\pm$0.07 & 1.751 & 0.1\\
$M_Z$ & Distance Residual (mag) & 9 & $n/a$ & 2.655 & 0.2\\
$B_{apparent}$ $-$ $V_{apparent}$ & Distance Residual (mag) & 4 & $-$0.51$\pm$0.42 & 2.723 & 0.1\\
$V_{apparent}$ $-$ $R_{apparent}$ & Distance Residual (mag) & 5 & $-$0.83$\pm$0.68 & 1.584 & $-0.8$\\
$R_{apparent}$ $-$ $I_{apparent}$ & Distance Residual (mag) & 13 & $-$0.28$\pm$0.51 & 1.916 & $-0.2$\\
$I_{apparent}$ $-$ $Z_{apparent}$ & Distance Residual (mag) & 8 & $-$0.04$\pm$0.11 & 2.490 & $-0.6$\\
GCD & $B_{apparent}$ $-$ $V_{apparent}$ & 5 & 0.08$\pm$0.10 & 2.222 & $-0.7$\\
GCD & $V_{apparent}$ $-$ $R_{apparent}$ & 5 & $n/a$ & 8.369 & 0.5\\
GCD & $R_{apparent}$ $-$ $I_{apparent}$ & 13 & $n/a$ & 10.663 & $-0.3$\\
GCD & $I_{apparent}$ $-$ $Z_{apparent}$ & 8 & 0.05$\pm$0.07 & 1.013 & 0.0\\
GCD & $A_V$ (mag) & $n/a$ & 0.02$\pm$0.03\tablenotemark{c} & $n/a$ & 0.1\\
\enddata
\tablenotetext{a}{Error ranges contain all linear fits with at least a 1\% chance of matching the distribution.  Values outside this range are ruled out at 99\% confidence.}
\tablenotetext{b}{We use $\rho$ as the symbol for the Spearman rank correlation coefficient.}
\tablenotetext{c}{Errors for $A_V$ vs. GCD were determined by a jack-knife test.}
\end{deluxetable}
\clearpage

\topmargin 0.0in
\newpage
\clearpage

\begin{figure}
\centerline{\psfig{file=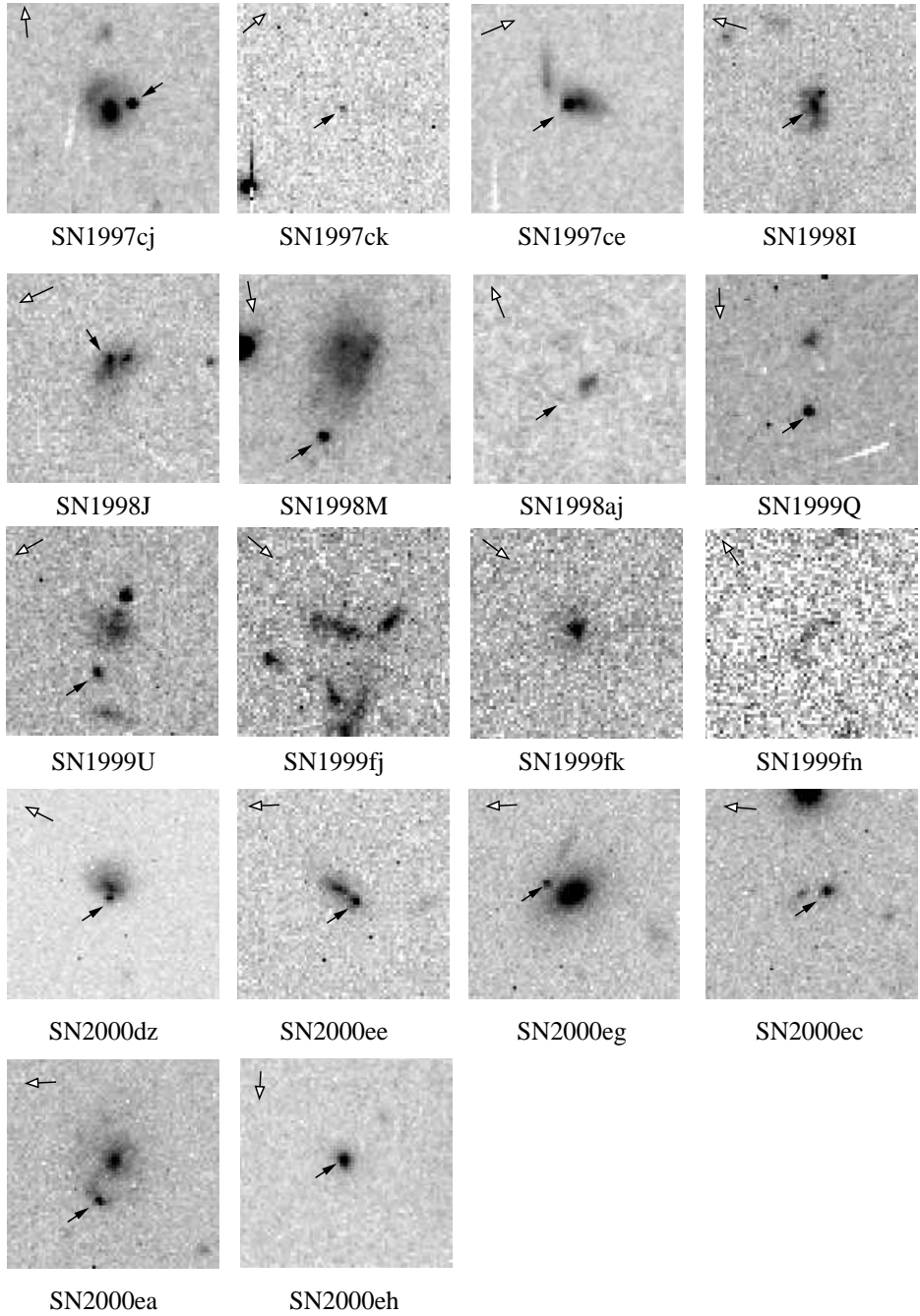,height=7.0in,angle=0}} 
\caption{Final images ($7''$ by $7''$) of the high-$z$ SN Ia host
galaxies.  In cases where only template images were available, no SN
is visible in the image.  In all other cases, the SN is indicated with
a filled arrow.  Open arrows in the upper-left corner of each image
point north; east is 90 degrees counterclockwise from north.}
\end{figure} 

\clearpage

\begin{figure}
\centerline{\psfig{file=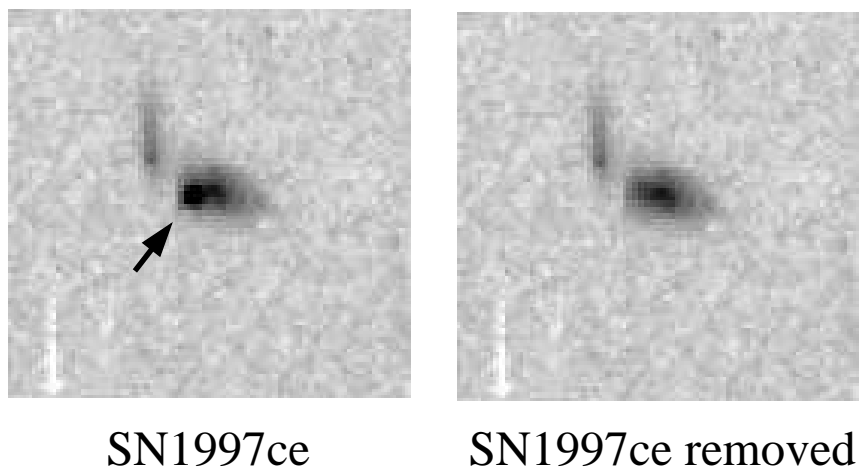,height=5.0in,angle=0}} 
\caption{An example of SN removal from a full image stack of
all epochs.  The left panel shows the $7'' \times 7''$ image of SN 1997ce in
the F814W filter before removal of the SN contamination.  The
right panel shows the same image after appropriate SN removal by linear
interpolation across the affected pixels.}
\end{figure} 

\begin{figure} 
\centerline{\psfig{file=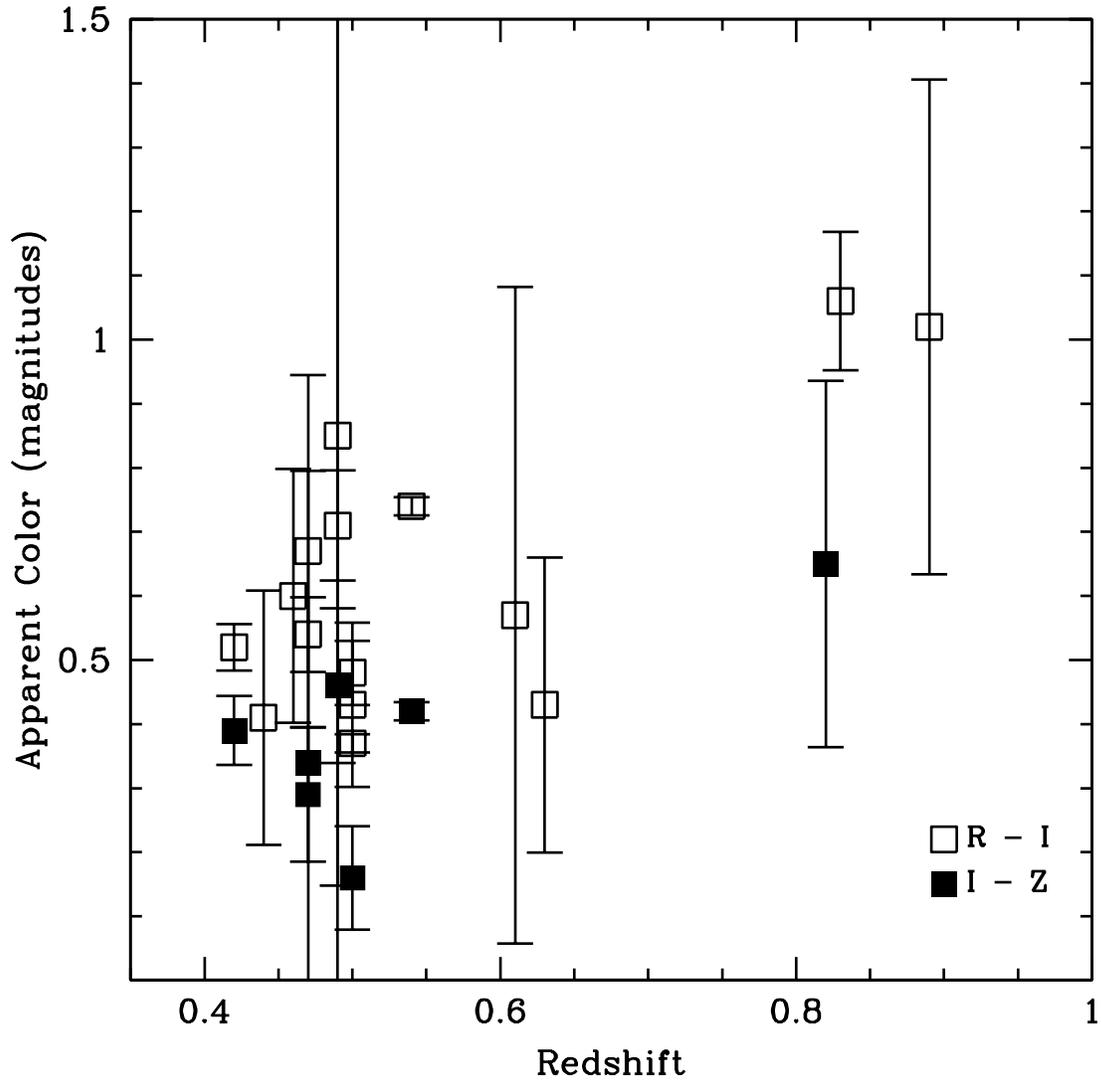,height=6.0in,angle=0}}
\caption{Plot of the apparent colors (in magnitudes) of the
host galaxies vs. their redshifts ($R - I$ = $m_{F675W} - m_{F814W}$,
$I - Z$ = $m_{F814W} - m_{F850LP}$).  The higher redshift galaxies are
indeed redder, revealing the effects of not performing K-corrections.}
\end{figure}

\begin{figure}
\centerline{\psfig{file=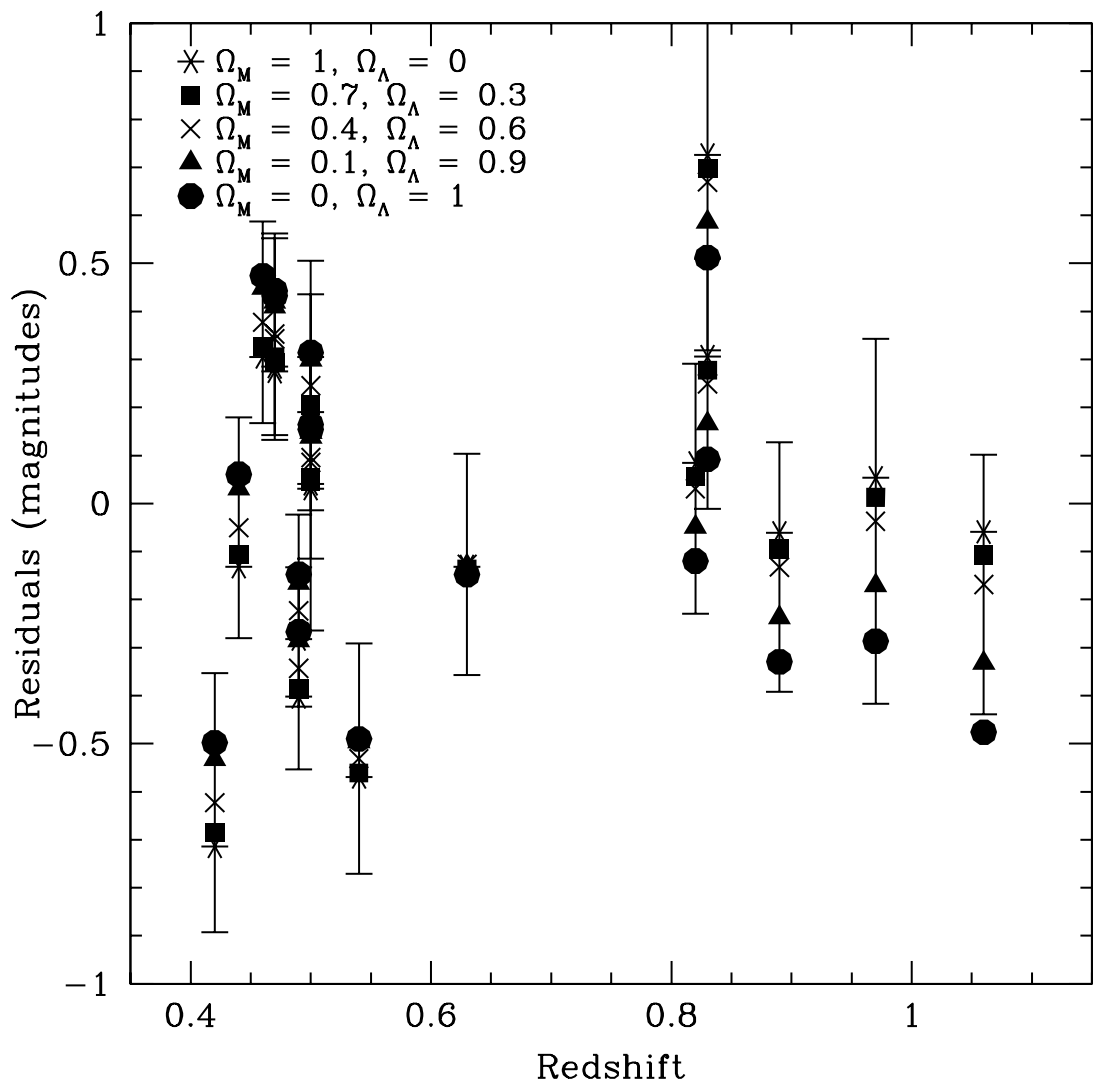,height=6.0in,angle=0}}
\caption{Distance modulus residuals (magnitudes) calculated for each
galaxy for several different choices of cosmological parameters.  For
reference, distance errors are shown only for the $\Omega_M = 0.4$,
$\Omega_{\Lambda} = 0.6$ case.  The distance errors are always larger
than the spread in residuals from the choice of cosmological
parameters.  As expected, there is no correlation of residual with
redshift for any of these parameter sets.}
\end{figure}

\begin{figure}
\centerline{\psfig{file=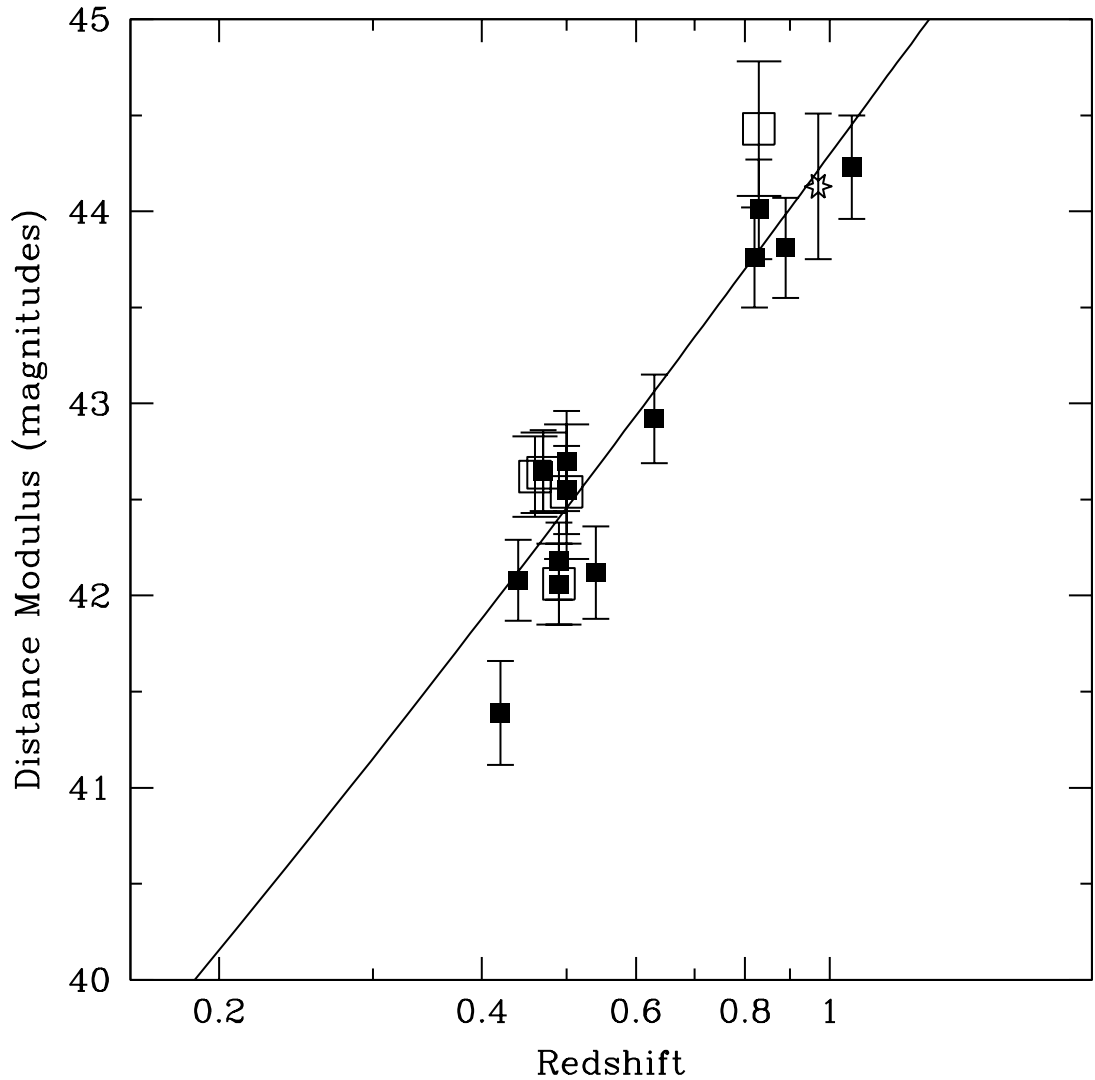,height=6.0in,angle=0}}
\caption{Best-fit Hubble diagram to our sample for an $\Omega_M =
  0.3,\ \Omega_{\Lambda} = 0.7$ cosmology.  For the zero-point adopted
  in this fit, $H_0=64$ km s$^{-1}$ Mpc$^{-1}$.  Events with
  early-type hosts are marked by open squares, those with late-type
  hosts by closed squares, those with no host by open stars.
  Demographic correlations in Figures 9-12 probe whether the
  deviations of points above and below this fit correlate with simple
  proxy indicators of progenitor properties.}
\end{figure}

\begin{figure}
\centerline{\psfig{file=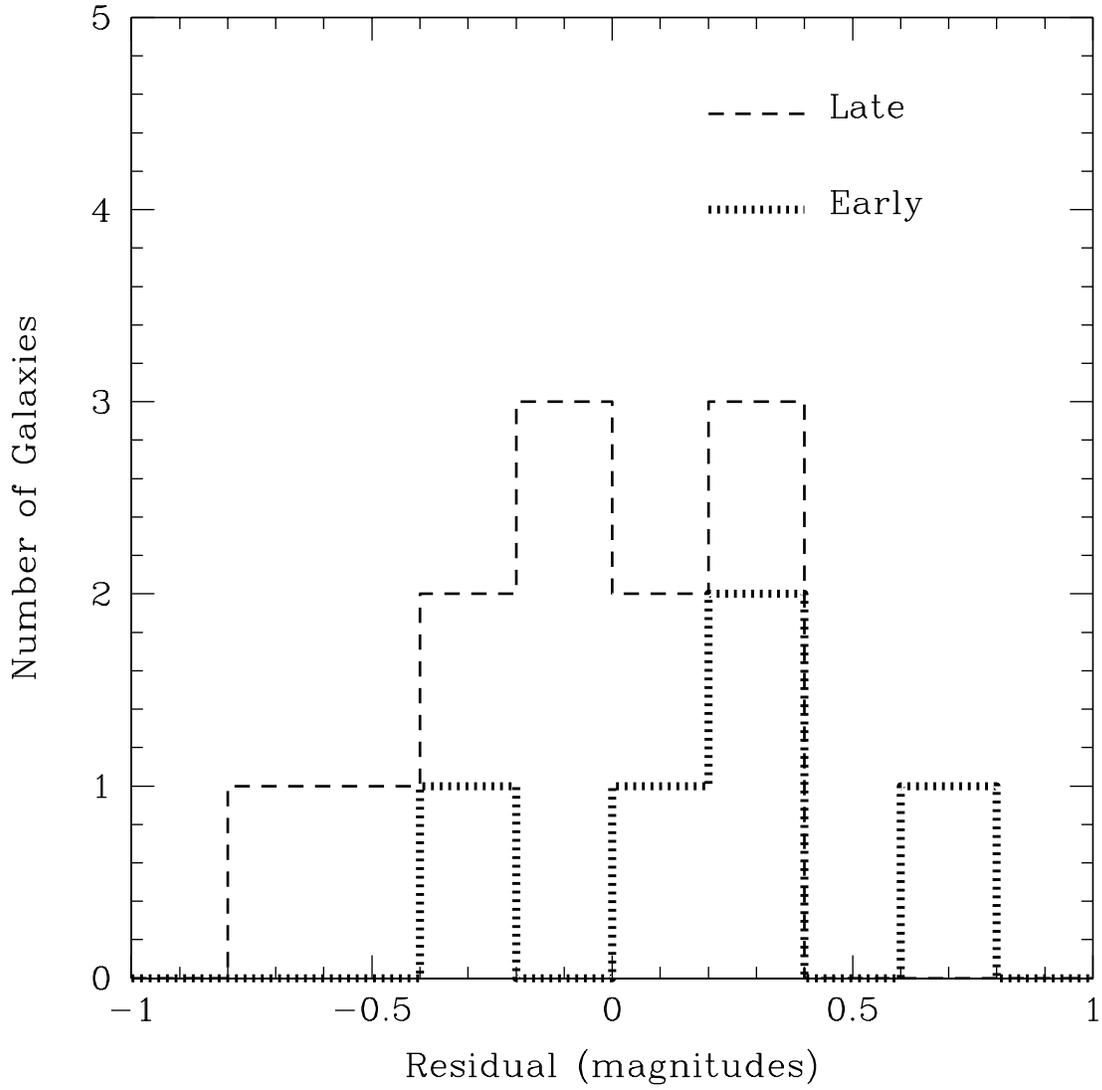,height=6.0in,angle=0}}
\caption{Distance modulus residuals calculated for each galaxy type
are shown in bins of 0.2 mag.}
\end{figure}

\begin{figure}
\centerline{\psfig{file=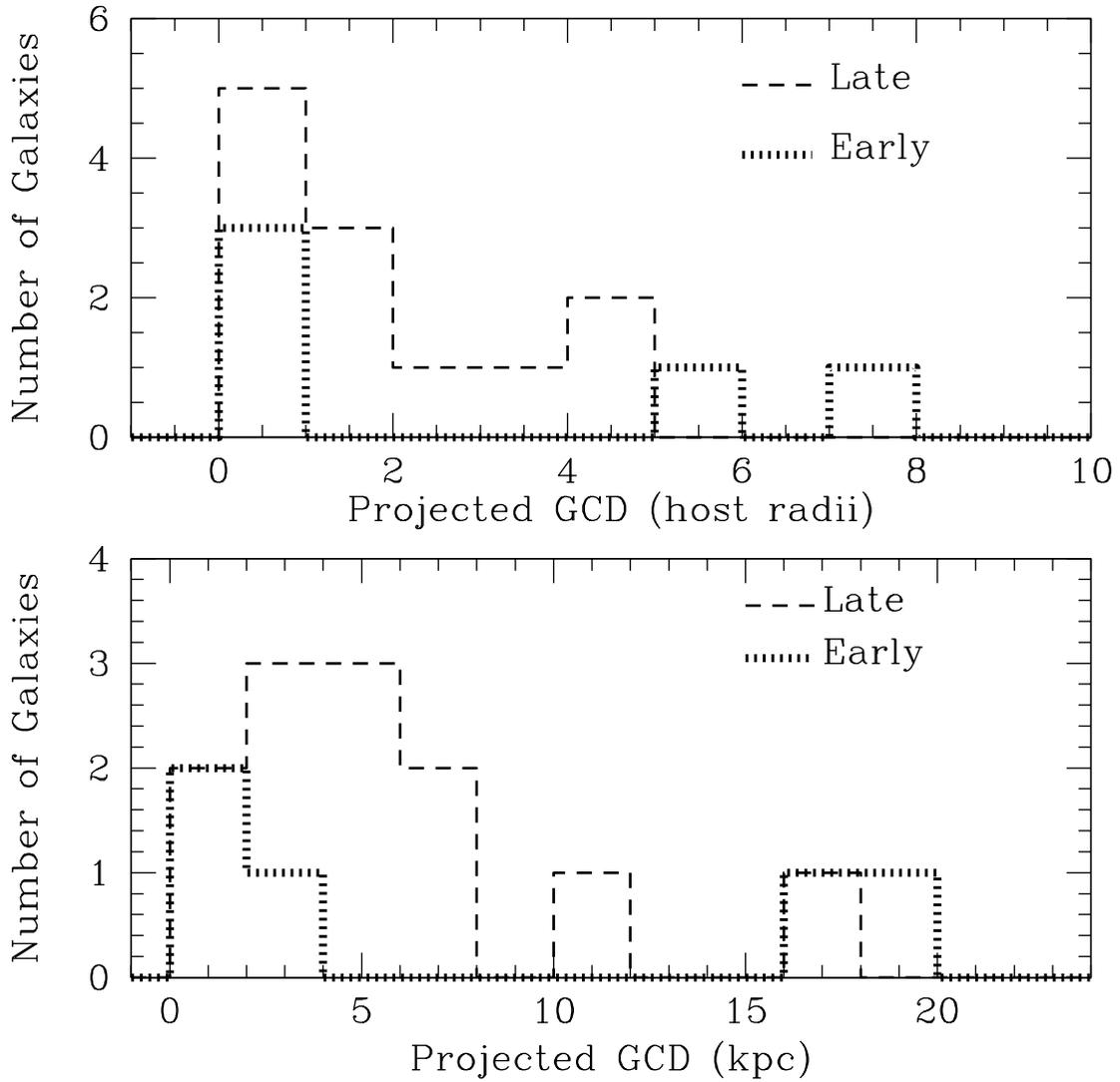,height=6.0in,angle=0}}
\caption{Histograms of the projected galactocentric distances of SNe
Ia measured for each galaxy type are shown in units of host scale
lengths ($top$) and kpc ($bottom$).}
\end{figure}

\begin{figure}
\centerline{\psfig{file=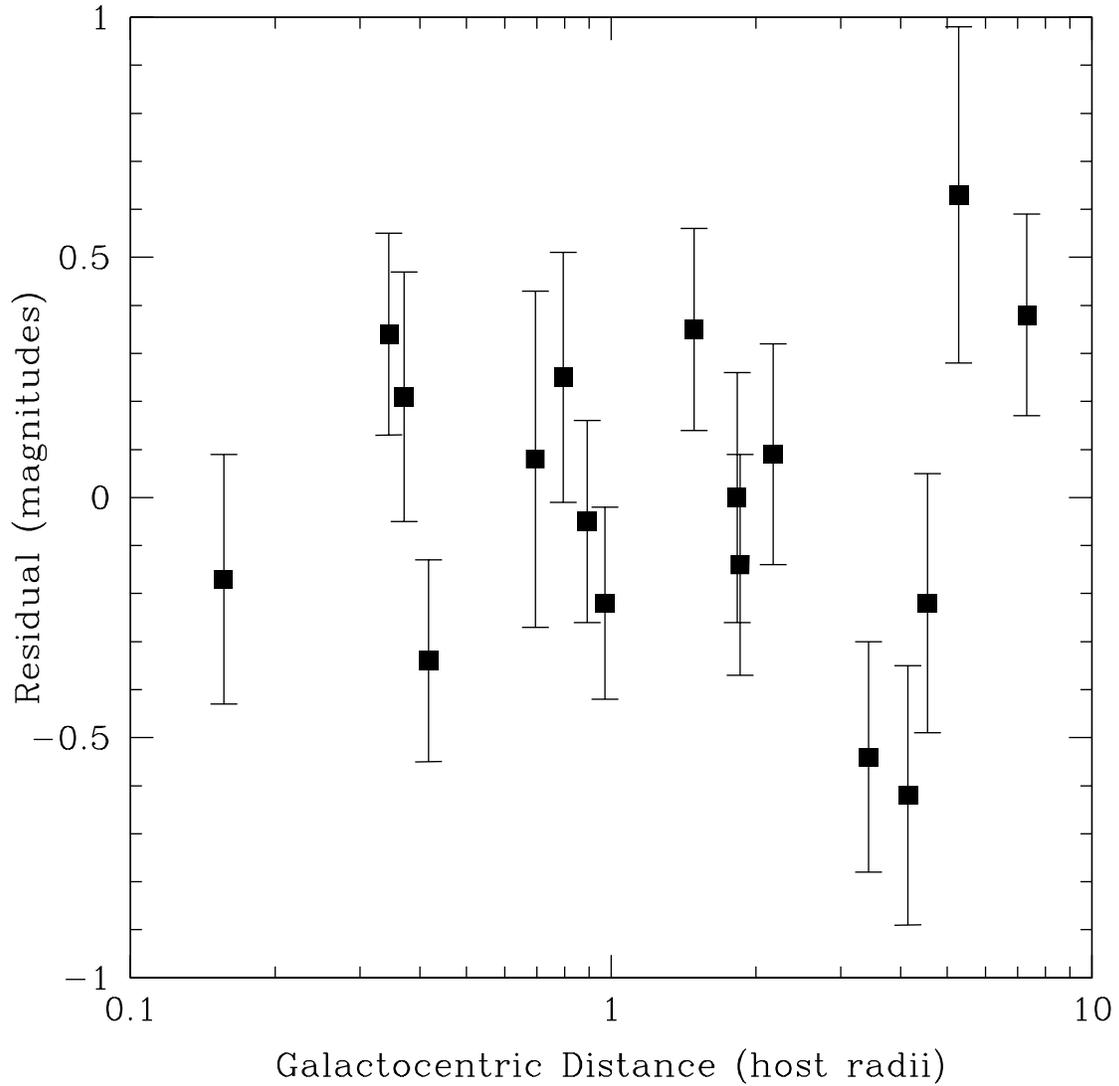,height=6.0in,angle=0}}
\caption{Plot of SN Ia galactocentric distance vs. the distance
modulus residuals of the Hubble fit for an $\Omega_M = 0.3,\
\Omega_{\Lambda} = 0.7$ cosmology.  A linear fit (see Table 3) confirms
that there are no significant correlations.}
\end{figure}

\begin{figure}
\centerline{\psfig{file=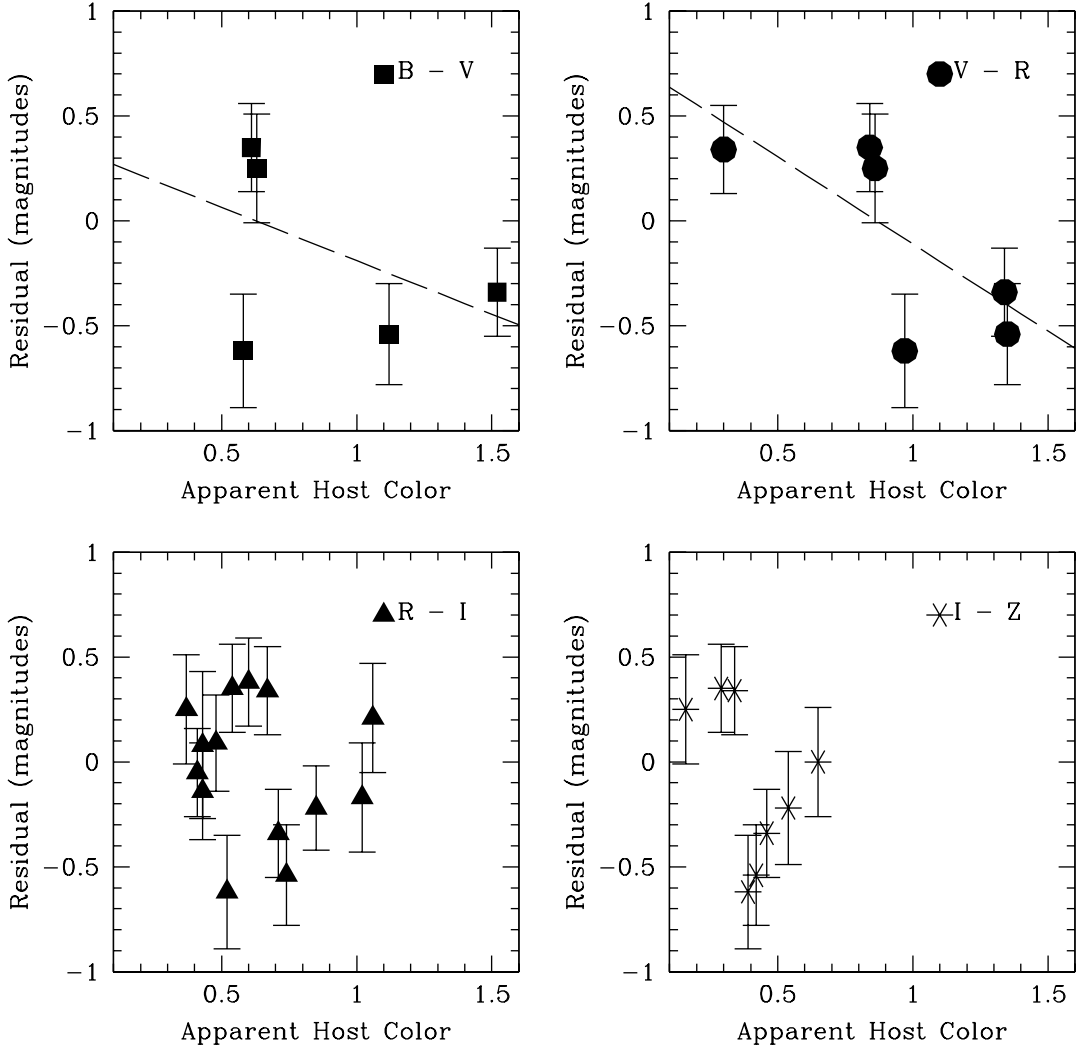,height=6.0in,angle=0}}
\caption{Plots of SN Ia host-galaxy apparent integrated colors ($B -
V$ = $m_{F450W} - m_{F555W}$, $V - R$ = $m_{F555W} - m_{F675W}$, $R -
I$ = $m_{F675W} - m_{F814W}$, $I - Z$ = $m_{F814W} - m_{F850LP}$)
vs. the distance modulus residuals of the Hubble fit for an $\Omega_M
= 0.3,\ \Omega_{\Lambda} = 0.7$ cosmology.  Linear fits suggest that
the residuals tend to be more negative for hosts redder in $B-V$ and
$V-R$.  Dashed lines show correlations for the best linear fits in
cases where the measured slope is not consistent with zero.}
\end{figure}

\begin{figure}
\centerline{\psfig{file=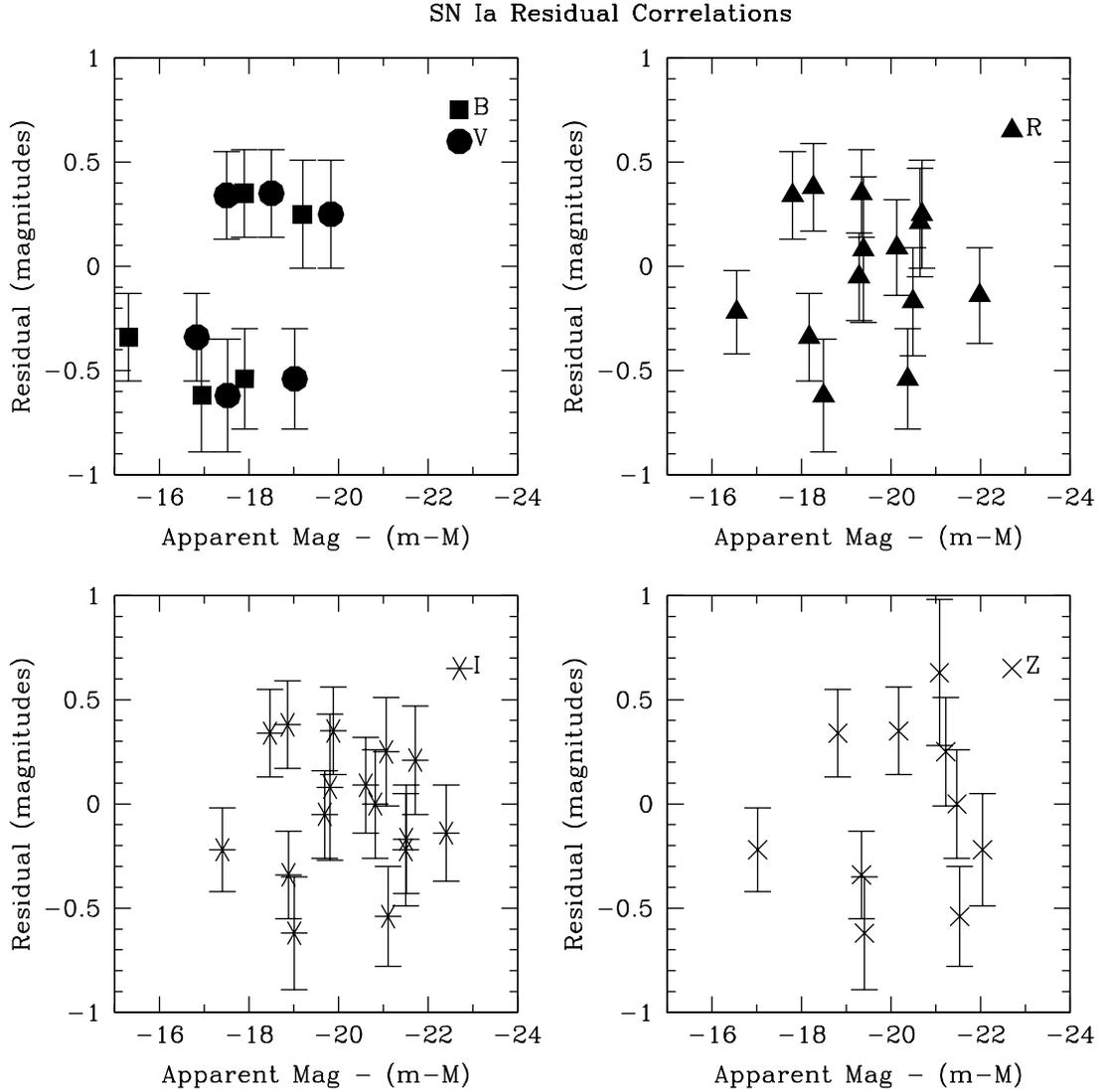,height=6.0in,angle=0}}
\caption{Plots of SN Ia host-galaxy $m_{apparent}$ minus distance modulus
vs. the distance modulus residuals of the Hubble fit for an $\Omega_M = 0.3,\
\Omega_{\Lambda} = 0.7$ cosmology.  Linear fits (see Table 3) confirm
no significant correlations, but the scatter is larger for the hosts
of the events observed in $B$ and $V$ (upper left panel), which were the
events in the year 2000.}
\end{figure}

\begin{figure}
\centerline{\psfig{file=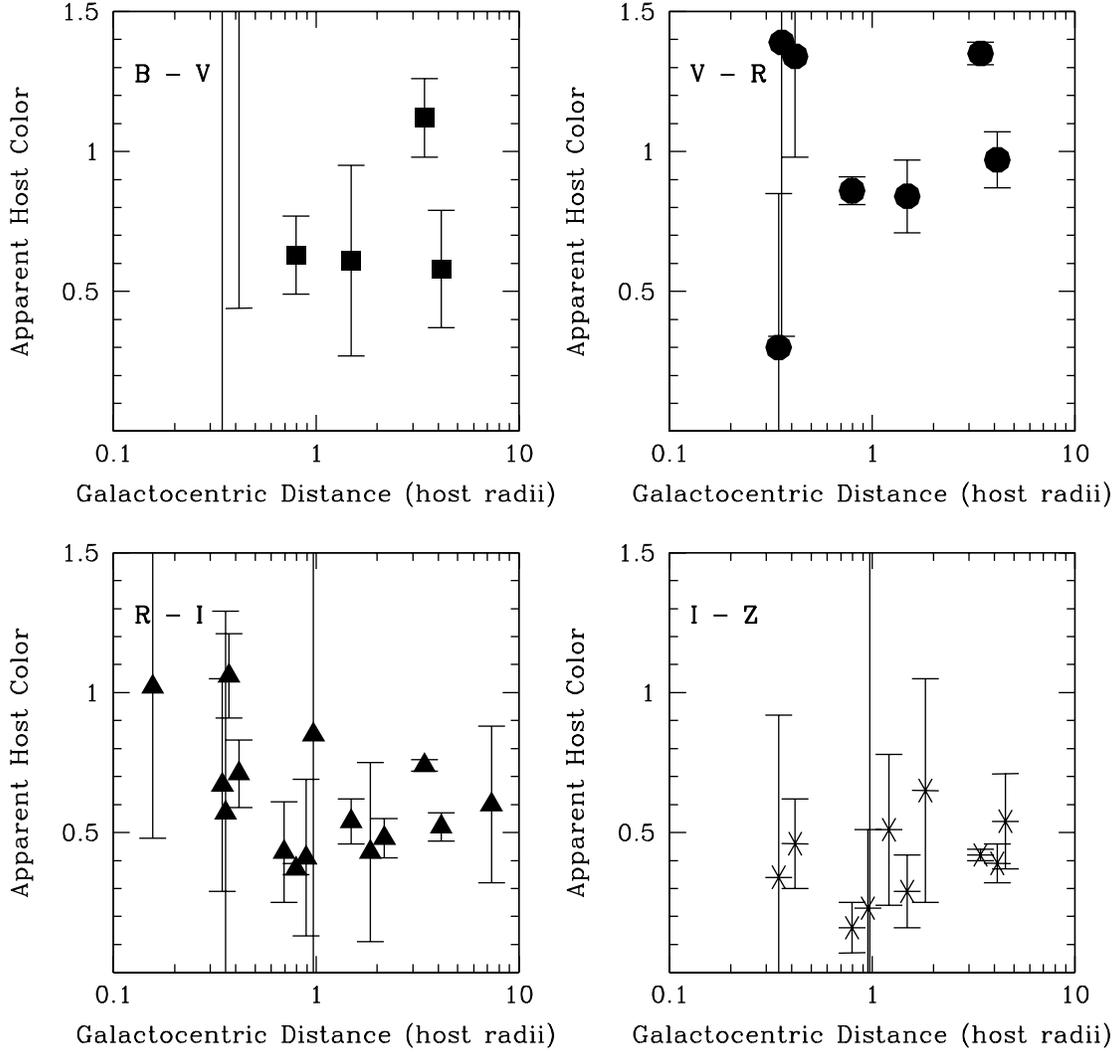,height=6.0in,angle=0}}
\caption{Plots of SN Ia galactocentric distance vs. the apparent
color of the host galaxy ($B - V$ = $m_{F450W} - m_{F555W}$, $V - R$ =
$m_{F555W} - m_{F675W}$, $R - I$ = $m_{F675W} - m_{F814W}$, $I - Z$ =
$m_{F814W} - m_{F850LP}$).  Linear fits (see Table 3) confirm no
significant correlations.}
\end{figure}

\begin{figure}
\centerline{\psfig{file=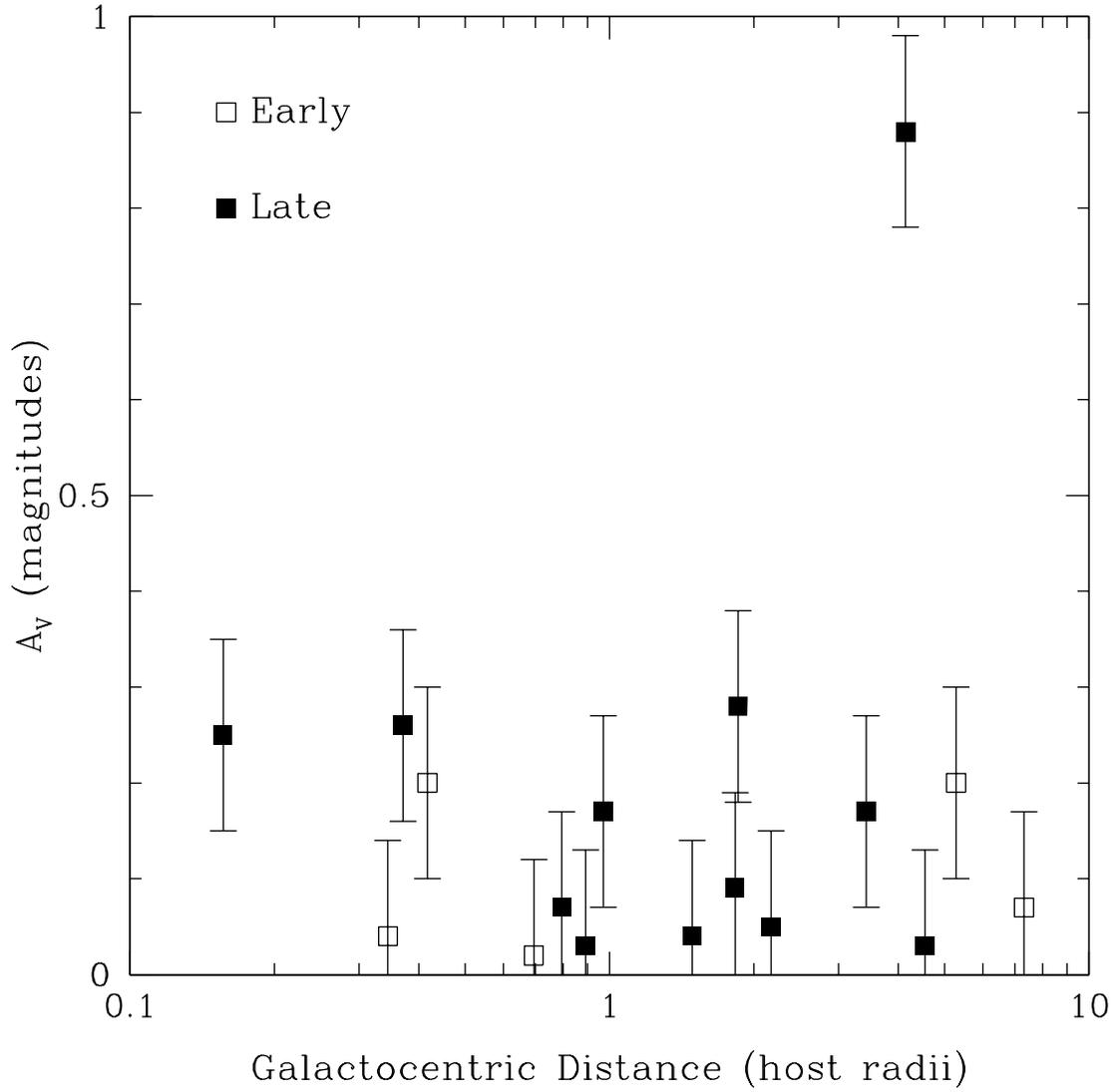,height=6.0in,angle=0}}
\caption{Plot of SN Ia galactocentric distance vs. $A_V$ measured by
\citet{tonry2003}.  Empty squares mark early hosts, and filled squares
mark late hosts.  A linear fit (see Table 3) confirms no significant
correlations.}
\end{figure}

\end{document}